\documentclass[a4paper,twocolumn,english,aps,accepted=2023-03-08]{quantumarticle}
\pdfoutput=1
\usepackage[utf8]{inputenc}
\usepackage[english]{babel}
\usepackage[T1]{fontenc}
\usepackage{amsmath}

\usepackage{hyperref}

\usepackage[numbers]{natbib}
\usepackage{mathtools}
\usepackage{amssymb}
\usepackage{bm}
\usepackage{bbold}
\usepackage{braket}
\usepackage[labelfont=bf]{caption}
\usepackage{graphicx}
\usepackage{float}
\usepackage{comment}
\usepackage{dcolumn}
\usepackage{multirow}

\usepackage{wrapfig}

\usepackage[all]{hypcap} 

\begin{document}

\title{Parity Quantum Optimization:~Encoding Constraints}

\author{Maike Drieb-Sch\"on}
\affiliation{Parity Quantum Computing GmbH, A-6020 Innsbruck, Austria}
\affiliation{Institute for Theoretical Physics, University of Innsbruck, A-6020 Innsbruck, Austria}

\author{Kilian Ender}
\affiliation{Parity Quantum Computing GmbH, A-6020 Innsbruck, Austria}
\affiliation{Institute for Theoretical Physics, University of Innsbruck, A-6020 Innsbruck, Austria}

\author{Younes Javanmard}
\affiliation{Parity Quantum Computing GmbH, A-6020 Innsbruck, Austria}

\author{Wolfgang Lechner}
\email{wolfgang@parityqc.com\\wolfgang.lechner@uibk.ac.at}
\affiliation{Parity Quantum Computing GmbH, A-6020 Innsbruck, Austria}
\affiliation{Institute for Theoretical Physics, University of Innsbruck, A-6020 Innsbruck, Austria}

\begin{abstract}
Constraints make hard optimization problems even more challenging to solve on quantum devices because they are typically implemented with large energy penalties and additional qubit overhead. The parity mapping, which has been introduced as an alternative to the spin encoding, translates the problem to a representation using parity variables that encode products of spin variables into a single parity variable. By combining exchange interaction and single spin flip terms in the parity representation, constraints on sums, products of arbitrary \textit{k}-body terms and sums over such products can be implemented without penalty terms in two-dimensional quantum systems.  
\end{abstract}
\maketitle

\section{Introduction}
\label{sec:introduction}
Constraints to optimization problems are crucial for many problems that are encountered in science, technology, and industry, ranging from scheduling problems to quantum chemistry~\cite{farhi_quantum_2001,allouche_computational_2014,GravelPRE2008,neukart_traffic_2017, rieffel_case_2015,hebrard_constraint_2010,johnson_quantum_2011_manufacturing_spins,cao_solving_2017}.
Quantum computing as a new paradigm of computing, which aims, among other things, at enhancing optimization algorithms by making use of quantum phenomena, may improve upon existing algorithms to solve these kinds of problems.
However, quantum computers are limited in coherence, control, and connectivity~\cite{arute2019quantum,bernien2017probing,KochPRAChargeInsensitive2007,saffman2010quantum,henriet2020quantum,bloch2008many} which makes encoding of optimization problems one of the current grand challenges in the field. Constraints are an additional complication to the encoding challenge and they are typically encoded via large energy penalties~\cite{bian_mapping_2016, douglass_constructing_2015, lucas2014ising} given as quadratic terms leading to fully connected interactions. These penalties introduce an additional energy scale and in most cases additional qubits and couplings to the computation, making algorithms, such as the quantum  approximate  optimization  algorithm (QAOA)~\cite{farhi2014quantum} or quantum annealing~\cite{nishimori98, ArnabDas_RMP2008, hauke_perspectives_2020}, less efficient.
Some recent works~\cite{a12040077, ItayHenPRApplied2016, ItayHenPRA2016} present more efficient quantum annealing methods to solve problems with linear equality constraints, which are pure sums over single spin variables,
while the penalty method is applicable for general equality and inequality constraints.
Reference~\cite{hadfield2019QAOA} gets rid of penalty terms in the QAOA by extending it to the quantum alternating operator ansatz, introducing a tailored mixing Hamiltonian for each constrained optimization problem specifically.

In this paper, we present the encoding of optimization problems with equality constraints in analog~\cite{ikeda_application_2019,irie_quantum_2019, stollenwerk_quantum_2020,HenYoung2011, ng_optimizing_2014, rieffel_parametrized_2014, venturelli_quantum_2016_chop_shop, rosenberg_solving_2016} and digital~\cite{hadfield2019QAOA, qaoa_wang, cook2020quantum} quantum computers using the parity quantum computing scheme~\cite{lechner2015quantum,compilerpaper,benchmarkpaper}, which is applicable to a broader range of equality constraints instead of pure sums over single logical qubits and only requires local interaction terms.

To encode constraints we introduce a combination of exchange interactions and spin-flip terms in combination with the parity encoding. The parity transformation encodes optimization problems in a lattice gauge model with local 3-body and 4-body interactions on a square lattice. We introduce exchange terms that only act on qubits that are part of the constraints and spin-flip terms that act on the rest of the qubits. Using a compiler~\cite{compilerpaper}, qubits can be arranged on the square lattice with flexibility. In particular, one can place qubits that are part of the constraint in close vicinity and thus allow for local hopping terms only. The points mentioned above make the method presented here implementable on current quantum devices for both quantum annealing and digital quantum computing.

We consider optimization problems in the form of a general $k$-body Hamiltonian
\begin{eqnarray}
\label{eq:H}
	H &=& \sum_i J_i s_i + \sum_{i<j} J_{ij} s_i s_j 
	+ \sum_{i<j<k} J_{ijk} s_i s_j s_k
	\\ \nonumber
	&+& \sum_{i<j<k<l} J_{ijkl} s_i s_j s_k s_l + ...,
\end{eqnarray}
where $s_i$ are the spin variables with values in $\{-1,1\}$. This Hamiltonian can be efficiently implemented using the parity transformation~\cite{compilerpaper}. We assume that the terms can go up to $N$-body terms and the $k$-body interaction tensors $J$ are sparse. We assume that the total number of qubits is $N$ and the number of terms in the Hamiltonian is $K$. 

\begin{figure}[t]
\centering
\includegraphics[width=1\columnwidth]{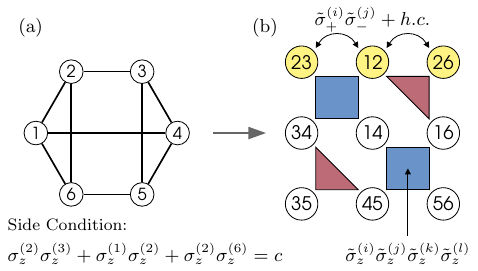}
\caption{\textbf{(a)} The standard approach to encode optimization problems on a quantum device is to separate the unconditioned from the conditioned part. The unconditioned part of the optimization problem is encoded as a spin model (top). Each node (circle) represents a logical spin, which can either be up or down. The edges (lines) represent interactions between the spins. The solution of the unconditioned part is the ground state of the spin model. In addition, a side condition on a subset of the spins is given (bottom).
\textbf{(b)} In the parity encoding, the problem is translated to a Hamiltonian that encodes the unconditioned as well as the conditioned part. The unconditioned part is encoded in a lattice gauge model, where nodes represent parity variables and interactions are local 3-body (red) or 4-body (blue) terms. The side conditions of the optimization problem are implemented as exchange terms among the conditioned parity qubits (yellow). In the parity encoding this specific side condition needs no extra resources.}
\label{fig1}
\end{figure}

In addition to the $k$-body optimization problem we consider a list of equality constraints where each constraint $m$ can be written as

\begin{align}
\label{eq:sumconst}
 c_m &= \sum^{N}_{i=1} g^{(m)}_i s_i + \sum^{N}_{i=1}\sum^{}_{j>i} g^{(m)}_{ij} s_i s_j\\ \nonumber
	 &+ \sum^{N}_{i=1}\sum^{}_{j>i}\sum^{}_{k>j}  g^{(m)}_{ijk} s_i s_j s_k + ... ~.
\end{align}
Equation~\eqref{eq:sumconst} includes sums and product terms, i.e., sums of single-body terms ${g^{(m)}_i s_i}$,
2-body terms ${g^{(m)}_{ij}}s_i s_j$,
3-body terms ${ g^{(m)}_{ijk} s_i s_j s_k}$,
and so on. The coefficients $g^{(m)}_i, g^{(m)}_{ij}, g^{(m)}_{ijk}, ...$ depend on the problem and its constraints.
Constraints of optimization problems comprising inequalities can be rewritten as constraints comprising only equalities by increasing the number of spins~\cite{lucas2014ising}.
Therefore, a large number of optimization problems with both equality and inequality constraints are covered by a set of equality constraints given by Eq.~\eqref{eq:sumconst}. 

Here we will restrict ourselves to constraints which are represented by Eq.~\eqref{eq:sumconst} and coefficients $g^{(m)}_{\alpha}$ which are zero or one. Furthermore, we only consider the case of disjoint constraints, which excludes all inequalities. Multiple constraints are disjoint if they have no common qubits, i.e., all qubits are only included in one of the constraints. Otherwise we call them overlapping constraints.
An idea of how to extend this method to overlapping constraints in the parity encoding and coefficients ${g^{(m)}_{\alpha}>1}$ (and therefore also inequalities) is illustrated and discussed in Appendix~\ref{inequalities}. However, since in these cases the implementation is unclear, it remains a task for future work.

\section{Encoding Constraints \label{encodingConstraints}}
The parity transformation translates each term in the Hamiltonian in Eq.~\eqref{eq:H} to a single parity qubit and all $k$-body interaction strengths $J_{i_1 \dots i_k}$ become local fields $J_{\alpha_i}$ only.
We denote parity qubits with $\tilde{\sigma}^{(i)}_{z}$ and their corresponding eigenstates with $\ket{\pm1}$. A detailed explanation of the parity transformation can be found in Ref.~\cite{compilerpaper} and in Appendix~\ref{appendix_parityTransormation},
which contains the transformation of the graph shown in Fig.~\ref{fig1} and some other examples of optimization problems in the presence of side conditions. In the following, we summarize the most important aspects for completeness: The logical qubits $\sigma^{(i)}_z$ with eigenvalues $\pm 1$ represent the spins $s_{i}$ in the $z$-basis. A parity qubit represents the product of $k$ logical qubits, e.g., $\tilde{\sigma}^{(1,2,3)}_{z} = \sigma^{(1)}_z\sigma^{(2)}_z\sigma^{(3)}_z$ for ${k=3}$.
The parity transformation transforms the graph into a parity layout, which then encodes the optimization problem in terms of parity variables.
This step increases the number of degrees of freedom from $N$ to $K$, where $N$ is the number of spins and $K$ is the number of non-zero terms in the problem Hamiltonian. If we have additional hard constraints as in Eq.~\eqref{eq:sumconst}, $K$ is the number of distinct spin products in both the non-zero terms of the problem Hamiltonian and the hard constraints combined plus the number of ancilla qubits, if they are needed.
Thus, $K-N+D$ parity constraints are introduced to restrict the low-energy subspace to the original size, where $D$ is related to the degeneracy of the Hamiltonian (see Ref.~\cite{compilerpaper} for further explanation). For the unconstrained optimization problem, the parity constraints are constructed from generalized closed cycles in the logical graph~\cite{compilerpaper}, for example $\tilde{\sigma}^{(1,2)}_{z} \tilde{\sigma}^{(2,6)}_{z}
\tilde{\sigma}^{(1,6)}_{z}$, which is a valid parity constraint as the corresponding product in the logical variables is always $+1$. The main challenge in the parity transformation is to select $K-N+D$ generalized closed cycles such that the resulting parity constraints can be laid out on a chip with local connectivity. For an all-to-all connected model with only 2-body terms, the choice of cycles and the layout was shown for the Lechner-Hauke-Zoller (LHZ) architecture~\cite{lechner2015quantum}. For other graphs and hypergraphs, the choice and arrangement of qubits is done by compilation~\cite{compilerpaper}.

To avoid confusion between parity constraints and constraints of the optimization problem, we will refer to the constraints of the optimization problem as side conditions.
Side conditions in the form of a single product of logical spins can be natively encoded in the parity mapping as shown in Ref.~\cite{compilerpaper}.
Side conditions in Eq.~\eqref{eq:sumconst} are arbitrary sums and products (polynomials) of logical qubits and will be discussed here.
Such side conditions can be transformed to the parity encoding by adding a physical qubit with zero local field for each product (including single-body terms) of logical qubits given in the side condition, which is not already in the Hamiltonian in Eq.~\eqref{eq:H}. This means that if each product of logical qubits in the side condition is already present in the problem Hamiltonian, then no additional physical qubits are needed. The example in Fig.~\ref{fig1} illustrates this case.

Traditionally, side conditions in optimization problems are
imposed with large energy penalties. For example, the side condition ${s_1s_2+s_2s_3 = +1}$ would be imposed by adding the energy term ${C_\textrm{penalty} (\sigma_{z}^{(1)}\sigma_{z}^{(2)} + \sigma_{z}^{(3)}\sigma_{z}^{(4)} - 1)^2}$ to the problem Hamiltonian. When $C_\textrm{penalty}$ is large enough, the ground state of the Hamiltonian satisfies the side condition. The challenges that come with this method are twofold: Firstly, the additional terms are higher-order interactions which introduces an additional overhead when embedding it in physical hardware with pair interactions~\cite{choi_minor-embedding_2008, VinciPRA2015,ItayHenPRApplied2016, ItayHenPRA2016}. Moreover, if we have optimization problems with a sparse problem Hamiltonian, side conditions can lead to a highly connected Hamiltonian leading to an overhead in qubits and couplings when embedding the side condition in currently available quantum hardware. Secondly, the large penalty strength $C_\textrm{penalty}$ introduces an additional energy scale to the system, which is unfavorable for the implementation as well as for the dynamics.

In this work, we use a different approach where side conditions are implemented in the dynamics rather than in the energy term, i.e., by using exchange terms as proposed in Refs.~\cite{ItayHenPRApplied2016, ItayHenPRA2016}. There, only side conditions which are sums over single logical qubits, can be realized by the given driver Hamiltonian. Furthermore, the authors note that in some of the investigated cases the experimental
feasibility of the driver Hamiltonians requires further consideration.
In contrast to that, we are able to use a similar driver Hamiltonian for side conditions, which are polynomial disjoint equality constraints (i.e., sums over products of logical qubits). In combination with a compiler~\cite{compilerpaper} that is able to lay out qubits, these exchange terms can be realized among nearest neighbors only, which makes it experimental feasible.

Our strategy is to transform the problem into the parity encoding, in which we can split up the parity qubits into two sets: 1.) qubits that are not part of the side condition terms $[\mathcal{U}]$, and 2.) qubits that are part of the side condition terms $[\mathcal{C}]$. Figure~\ref{fig1} illustrates this setup for a particular example. In this example, the side condition includes the physical qubits $\tilde{\sigma}_{z}^{(2,3)}$, $\tilde{\sigma}_{z}^{(1,2)}$, and $\tilde{\sigma}_{z}^{(2,6)}$ which are in $[\mathcal{C}]$ (yellow) and the rest of the qubits are in $[\mathcal{U}]$ (white).
If new physical qubits are required for the side condition, generally one has to add parity constraints to integrate the physical qubits into the physical lattice of the parity encoding.
In this example we do not need to add any physical qubits or any parity constraints for the side condition, because each product of logical qubits present in the side condition
($\sigma_{z}^{(2)}\sigma_{z}^{(3)}$, $\sigma_{z}^{(1)}\sigma_{z}^{(2)}$ and $\sigma_{z}^{(2)}\sigma_{z}^{(6)}$) is already present in the problem Hamiltonian. For a side condition of the form $c= \sigma_{z}^{(1)}\sigma_{z}^{(2)}\sigma_{z}^{(3)} + \sigma_{z}^{(2)}\sigma_{z}^{(3)}$, one would have to add the two physical qubits $\tilde{\sigma}_{z}^{(1,2,3)}$ (for the side condition) and $\tilde{\sigma}_{z}^{(2)}$ (ancilla qubit) and the parity constraint $\tilde{\sigma}_{z}^{(2,3)}\tilde{\sigma}_{z}^{(1,2)}\tilde{\sigma}_{z}^{(1,2,3)}\tilde{\sigma}_{z}^{(2)}$. For further details on the interplay between side conditions and the parity encoding, including decoding strategies with and with out side condition and in the presence and absence of errors we refer to Appendix~\ref{appendix_parityTransormation}.

In the annealing protocol, which is introduced below, terms that are part of $[\mathcal{U}]$ are driven with single body terms $\tilde{\sigma}_{x}$, while terms in $[\mathcal{C}]$ are driven by exchange terms. These exchange terms (or hopping terms) between qubits in the set $[\mathcal{C}]$ are of the form  $\tilde{\sigma}^{(k_i)}_+ \tilde{\sigma}^{(k_j)}_- + \tilde{\sigma}^{(k_i)}_- \tilde{\sigma}^{(k_j)}_+ $, where $\tilde{\sigma}^{(j)}_{\pm} = \tilde{\sigma}^{(j)}_x \pm i \tilde{\sigma}^{(j)}_y$. The global ground state of this exchange term is in the zero magnetization sector. However, the magnetization sector required by a side condition is $\sum^{n}_{i=1} \tilde{\sigma}_{z}^{(k_i)} = c$. The exchange Hamiltonian preserves the magnetization required by the side condition during the annealing process. The final state is the lowest-energy state in the magnetization (symmetry) sector which fulfills the side condition -- which may or may not be the unconstrained ground state. By starting in a configuration that satisfies the side condition, as the exchange term conserves the sum, the state will stay in the side-condition-satisfying subspace, i.e., the subspace of $H(t)$ that satisfies the side conditions. The parity constraints (i.e., 4-body or 3-body terms) act on all qubits, including $[\mathcal{C}]$ and $[\mathcal{U}]$, the hopping term only on qubits in $[\mathcal{C}]$ and the single spin flip term $\tilde{\sigma}_{x}$ only on qubits in $[\mathcal{U}]$.\\

In the following we will describe the protocols for Hamiltonian dynamics and digital quantum computing.

\section{Hamiltonian Dynamics}
\label{sec:hamiltonDynamics}

The optimization problem with side conditions encoded as described above can be solved using an adiabatic quantum computing protocol. Our protocol introduces a few novelties compared to standard quantum annealing: 1.) The initial state is a classical state similar to reverse annealing protocols~\cite{YamashiroPRA2019}, which is chosen such that the side condition is fulfilled. 2.) We introduce exchange terms acting on parts of the system and single spin flips acting on the rest.
3.) The target state is not necessary the ground state but we instead aim at obtaining a lowest-energy state in a different symmetry sector.
With the prescriptions 1.)\,-\,3.), the annealing process only populates states that fulfill the side condition, which is ensured by the driver Hamiltonian.
It is important that the driver Hamiltonian reaches each side-condition-fulfilling eigenstate, but no others.

As in all adiabatic protocols, we start from the ground state of an initial Hamiltonian. In quantum annealing, this is the ground state of the driver Hamiltonian.
In the following we consider the case of a single side condition.
This protocol can be easily extended to multiple disjoint side conditions, although generally ancilla qubits and extra couplings for a suitable parity layout might be required. We refer to Appendix~\ref{appendix_multilDisjiontDynamics} for more details.
The initial Hamiltonian reads as follows:
\begin{equation}
	H_{\textrm{init}} = \sum_i^{[\mathcal{U}]} \varepsilon_i \tilde{\sigma}^{(i)}_{z} + \sum_i^{[\mathcal{C}]} \eta_i \tilde{\sigma}^{(i)}_{z}.
	\label{eq:Hinit}
\end{equation}
The sum $[\mathcal{C}]$ runs over all qubits that are included in the side condition, which we call conditioned qubits in the following.
The sum $[\mathcal{U}]$ runs over the rest of the qubits.
The coefficients $\varepsilon_i$ are chosen randomly from $\{-1,1\}$. The coefficients $\eta_i$ are chosen from $\{-1,1\}$ such that the side conditions are satisfied, e.g., for a side condition with $\sum_i^{[\mathcal{C}]}\tilde{\sigma}^{(i)}_{z} = +c$, the local fields are chosen to fulfill $\sum_i^{[\mathcal{C}]}\eta_i=-\langle c\rangle_\textrm{target}$, where $\langle c\rangle_\textrm{target}$ is the target expectation value of the side condition. 
The optimization problem is encoded in the local fields of the final Hamiltonian consisting of 
\begin{equation}
\label{eq:localH}
	H_{\textrm{final}} = \sum_i^K J_{i} \tilde{\sigma}^{(i)}_{z}
\end{equation}
and
\begin{equation}
\label{eq:parityH}
	H_{\textrm{C}} = \sum_{l=1}^{P} C_l \prod_{i}^{[P_l]} \tilde{\sigma}^{({i_l})}_{z}. 
\end{equation}
Here, $C_l$ is some coupling strength for the $l$-th parity constraint $P_l$ and the sum runs over all $P$ parity constraints $P_l$, where $P$ is $K-N+D$ and the product $[P_l]$ runs over all qubits in the parity constraint $P_l$. Note that the number of qubits in $P_l$ is $|[P_l]| = k$ for a $k$-body parity constraint. In Fig.~\ref{fig1}, we have two 3-body and two $4$-body parity constraints.
The coupling strength $C_l$ is chosen to be sufficiently large, such that all parity constraints $\prod_{i}^{[P_l]} \tilde{\sigma}^{({i_l})}_{z}$ have eigenvalue $+1$ for all $l = 1, ..., P$, if the system is in a physical ground state of $H_{\textrm{final}}+H_{\textrm{C}}$.

The single spin flip driver Hamiltonian acts only on qubits
not associated with the side condition terms,
\begin{equation}
	H_{\textrm{x}} = \sum_i^{[\mathcal{U}]} \tilde{\sigma}^{(i)}_{x}.
\end{equation}
In addition to the terms above, an exchange term is introduced among $[\mathcal{C}]$, which is,
\begin{equation}
\label{eq:Hexchange}
	H_{\textrm{exch.}} = \gamma_e \sum_{\langle i,j \rangle}^{[\mathcal{C}]^*} \tilde{\sigma}^{(i)}_+ \tilde{\sigma}^{(j)}_-  +  h.c.
\end{equation}
Here, the bracket $[]^*$ denotes a sum over all pairs $\langle i, j \rangle$ of neighboring qubits connected via an exchange term in $[\mathcal{C}]$  e.g., in Fig.~\ref{fig1} the number of terms in the side condition is $|[\mathcal{C}]|=3$ and the number of exchange terms $|[\mathcal{C}]^*|=2$ as there are two possible exchange moves that can be done with nearest neighbors.
The full time-dependent protocol reads
\begin{eqnarray}
\label{eq:Ht}
	H(t) = & A(t) & H_{\textrm{init}} + B(t) H_{\textrm{x}} + C(t) H_{\textrm{C}} \\ \nonumber
	+ & D(t)&  H_{\textrm{final}} + E(t) H_{\textrm{exch.}}.
\end{eqnarray}
Here, the functions A,B,C,D, and E are chosen as follows  
\begin{center}
\begin{tabular}{ c c c c c}
  & initial~ & ~~during~~ & ~final \\ 
 A & 1 & finite & 0 &\\ 
 B & 0 & finite & 0 &\\  
 C & 0 & finite & 1 &\\  
 D & 0 & finite & 1 &\\  
 E & 0 & finite & 0 &.
\end{tabular}
\end{center}
There exist other works, which use protocols that differ from the typically used time-dependent Hamiltonian ${H(s) = (1-s)H_{\textrm{init}} + s H_{\textrm{problem}}}$ for quantum annealing~\cite{albashLidar, hauke_perspectives_2020}.
Some of them are similar to the protocol presented here in that they have an additional time-dependent term that is only switched on at intermediate times (i.e., zero for $s=0,1$ and finite else), but they have different intentions.
This includes works that stay in specific sub-spaces~\cite{gluedTreesProblem} or use randomly chosen catalyst Hamiltonians to improve the performance of the annealing process~\cite{catalys1Crosson, catalyst2Troyer}.
While most of them have three individually driven terms, the initial, the intermediate and the final one, there is the quantum counter-diabatic annealing used in Ref.~\cite{hartmann2019rapid} where each sub-term of these three Hamiltonian parts is driven individually in time.
Here we have an intermediate driver Hamiltonian, which is divided in two parts $H_{\textrm{x}}$ and $H_{\textrm{exch.}}$ where each part can be driven individually in time. We are not interested in simply modifying the traditional path by adding an intermediate Hamiltonian. Our intention is to drive the system only within a certain subspace of the full Hilbert space by starting in a randomly chosen classical state of this subspace and creating superpositions of all states spanning this subspace by switching on the intermediate driver Hamiltonian. The method presented here goes beyond other existing methods.  
The basic idea of the exchange Hamiltonian~\eqref{eq:Hexchange} introduced in Refs.~\cite{ItayHenPRApplied2016, ItayHenPRA2016} can only handle side conditions that are sums over single logical qubits. Indeed, due to the parity transformation, we can map sums over arbitrary products of logical qubits, i.e., polynomial side conditions, to sums over single physical qubits, such that we are also able to solve problems with polynomial side conditions. This generalization from linear to polynomial side conditions is one of the main results of this work.

The Hamiltonian~\eqref{eq:parityH} contains each physical qubit at least once and the sum runs over all parity constraints,
which ensures that we only get results with a valid correspondence to the logical problem.
The exchange Hamiltonian swaps only the spin direction of neighboring qubits, such that the sum over all side conditioned qubits stays constant. 
With a classical preprocessing step, side conditions which can not be fulfilled can be excluded (for a detailed discussion we refer to Appendix~\ref{appendix_InterplayParityConstraintsSideConditions}).
If a solution of the problem exists (and the parity-constraint strength is large enough), no parity constraint will be violated and all side conditions will be fulfilled in the end.
However, to ensure that the target ground state is reached, it could be necessary to set the parity constraint strengths to higher values.
In the case that side conditions are unsolvable the final states would always be parity-constraint-violating states.
For more details see Appendix~\ref{appendix_InterplayParityConstraintsSideConditions}.

Since the exchange Hamiltonian acts only on neighboring qubits, the parity transformation has to translate the logical graph and the product terms of the side conditions into a common parity layout where all side conditioned qubits are connected via adjacent conditioned qubits. If this is not possible with the resulting physical qubits, one has to introduce ancilla qubits. With these ancilla qubits one must construct new parity constraints, which make it possible to find a suitable layout.
Therefore, laying out the parity qubits for exchange terms among neighbors, just like adding product terms of the side condition, which are not included in the problem Hamiltonian, leads to extra qubits and therefore additional and possibly different parity constraints.
It is possible that the encoding of the side conditions does not need any extra qubits, as it is the case for the example presented here. 
For some special case it is even possible to subsume a side condition into a parity constraint or to drop a parity constraint due to the given side condition. There are more details on the usage of ancillas and these special cases in Ref.~\cite{compilerpaper} and in Appendix~\ref{appendix_parityTransormation}.

As mentioned above,
as long as in the parity encoding the multiple side conditions do not overlap, i.e., they are disjoint, the method presented here can be extended to cases with multiple side conditions.
Side conditions overlap in the parity encoding only if different side conditions have the same product terms of logical qubits.
We analyse these cases in in Appendices~\ref{appendix_parityTransormation} and \ref{appendix_multilDisjiontDynamics}.

\subsection{Numerical Simulations}
\label{subsec_numericalsimulations}

\newcommand{\ugs}{\Psi^{(u)}_{gs}}
\newcommand{\cgs}{\Psi^{(c)}_{gs}}
\newcommand{\opH}{H_{\textrm{problem}}}
\newcommand{\inits}{\Psi(0)}

To interpret the results we have to distinguish between the ground state of the optimization problem Hamiltonian without side conditions,
${H_{\textrm{problem}} = H_{\textrm{final}} + H_{\textrm{C}}}$, which we call unconditioned ground state  $\ugs$ and the ground state of the optimization problem with side condition, i.e., the side-condition-satisfying ground states $\cgs$.
The state $\cgs$ is what we are seeking and corresponds to the lowest-lying eigenstate of $H_{\textrm{problem}}$, which satisfies the side conditions.

In the following, we will give an example for (i) the case $\ugs = \cgs$ and (ii) the case $\ugs \neq \cgs$.
In the latter case a crossing of the energy levels must occur. We will present the protocol of the annealing process, discuss  diabatic transitions and consider $\cgs$ and/or $\ugs$ to be degenerate.
We will also consider the role of the initial state.
The numerical results are presented for a random example where the unconditioned optimization problem is encoded together with the side condition in the parity model with 9 qubits, as shown in Fig.~\ref{fig1}{\color{quantumviolet}(b)}.
For details on the annealing protocol of all examples, we refer to Appendix~\ref{appendix_annealingprotocol}.
The random problem instance is encoded in the local fields $J_{i}$ as the field vector
\begin{equation}
    \label{loc_fields_main}
    J = (0.8, ~0.6, ~1.0, ~1.0, ~0.7, ~0.7, ~0.1, ~0.6, ~0.8)^T.
\end{equation}
For better comprehensibility, the first ten eigenstates of this problem are explicitly written out in
Eq.~\eqref{compiledSCExample} of Appendix~\ref{appendix_multilevelcrossing}.
The components of the field vector and of the system states correspond to the qubits in Fig.~\ref{fig1} in the order from top to bottom and from left to right, i.e., as $(23, 12, 26, 34, 14, 16, 35, 45, 56)^T$. 
\begin{figure}[!t]
\centering
\includegraphics[width=\columnwidth]{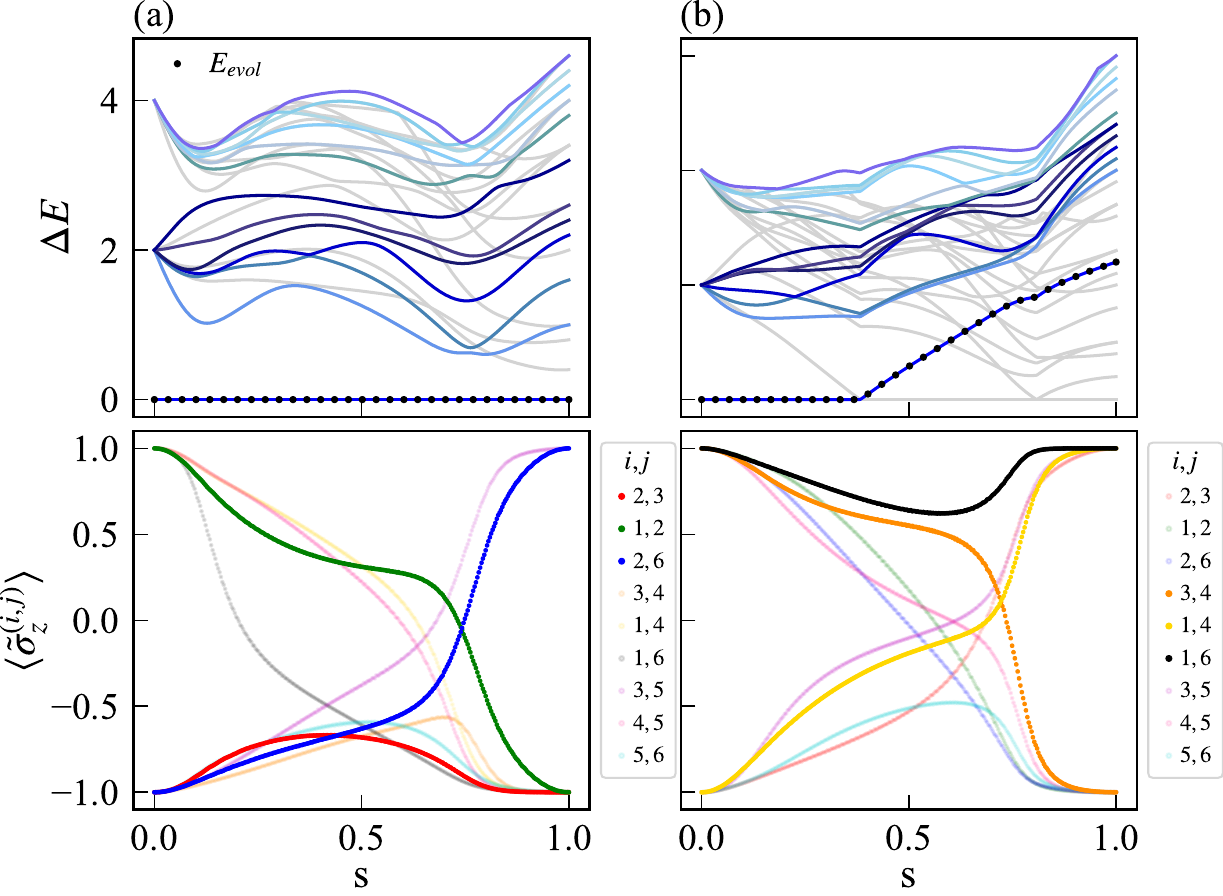}
\caption{Adiabatic sweep for a system with $N=9$ qubits and time $s(t)=t/t_f$ with $t_f = 500$.
The top row in Fig.2~(a) and (b) shows the instantaneous energy spectrum, where all instantaneous eigenenergies of the time-dependent Hamiltonian in Eq.~\eqref{eq:Ht} are given with respect to the ground state energy, $\Delta E = E_i - E_0$.
The energies of conditioned eigenstates, which lie in the side-condition-satisfying subspace are colored, while all the other ones are grey.
The evolution of the system state energy $E_{\rm{evol}}$ (black dots) with time follows the instantaneous energy levels of the time-dependent Hamiltonian. The bottom row of Fig.2~(a) and (b) shows the time-dependent expectation value ${\langle \tilde{\sigma}_z^{(i)} (s)\rangle}$ for each spin $i$, following the systems configuration state evolution in time.
For the spins included in the side condition, the expectation values are highlighted, while the unconditioned ones are presented in pale colours.
\textbf{(a)}~The side condition is given by the sum constraint
${\langle\tilde{\sigma}_{z}^{(2,3)} + \tilde{\sigma}_{z}^{(1,2)} + \tilde{\sigma}_{z}^{(2,6)}\rangle = -1}$ and
the initial state is set to
${\ket{\Psi_{(a)}(0)} = \ket{-1, 1, -1, -1, 1, 1, -1, 1, -1}}$. The initial state and the unconditioned ground state $\ugs$ satisfy the side condition.
The energy of the system follows the instantaneous ground state energy.
\textbf{(b)}~For the side condition given by ${\langle\tilde{\sigma}_{z}^{(3,4)} + \tilde{\sigma}_{z}^{(1,4)} + \tilde{\sigma}_{z}^{(1,6)} \rangle = 1}$,  the conditioned ground state $\cgs$ is equal to the eighth excited state of the unconditioned problem. 
The initial state is set to ${\ket{\Psi_{(b)}(0)} = \ket{-1, 1, 1, 1, -1, 1, -1, 1, -1}}$, which satisfies the side condition. Therefore, the energy of the time evolved system state crosses multiple energy levels of the Hamiltonian in Eq.~\eqref{eq:Ht}. The \textit{kinks} in the energy spectra (as the one near $s=0.4$ and $s=0.8$ in (b)) occur when higher-energy states \textit{cross} the ground state because the energy difference to the instantaneous ground state is shown.
These \textit{crossings} of the instantaneous ground state with states of different magnetization sectors are allowed and even necessary here.
}
\label{fig2}
\end{figure}
As the initial state satisfies the side condition
$c = \sum_{i}^{[\mathcal{C}]} \tilde{\sigma}_{z}^{(i)}$, the system ends up in the conditioned ground state.
Nevertheless, the instantaneous energy spectrum and the dynamics of the evolution of the system are affected by the choice of the initial state. Therefore, for some choices of initial states, \textit{multiple level crossings} to instantaneous excited energy states and back to the instantaneous ground state of the time-dependent Hamiltonian given in Eq.~\eqref{eq:Ht} are possible (even for $\ugs = \cgs$, further discussed in Appendix~\ref{appendix_multilevelcrossing}) while in other cases the system remains in the instantaneous ground state during the whole annealing process, as in Fig.~\ref{fig2}{\color{quantumviolet}(a)}.
Furthermore, the choice of the initial state affects the minimal energy gap between the conditioned ground state and the first excited state inside the side-condition-fulfilling subspace during the annealing process.
This minimum energy gap is crucial for the adiabatic condition~\cite{PhysRevLett.102.220401} and gives a lower limit for the computation time $t_f$.
We remark that by increasing the constant prefactor $\Gamma$ of the driver Hamiltonian we are able to shift the required final annealing time to lower values. In Appendix~\ref{appendix_diabatic} we discuss this in more detail and present an example for the adiabatic condition of the given annealing protocol.

If $\ugs \neq \cgs$, the final state $\Psi(t_f)= \cgs$ is equal to the lowest-energy eigenstate that satisfies the side condition. Therefore, there must be a level crossing, such that system can end up in the lowest-energy state in the symmetry sector satisfying the side condition.
In Fig.~\ref{fig2}{\color{quantumviolet}(b)} we consider the same problem example, and choose the side conditions such that the solution to the problem without side condition is different to the one with side condition.

Note that the energy of the instantaneous system state \textit{crosses} different energy levels of the spectrum of the time-dependent Hamiltonian given in Eq.~\eqref{eq:Ht}.
If the solution of the problem with side condition is the $n$-th excited eigenstate of the problem without side condition, the instantaneous energy of the system has to \textit{cross} at least $n$ energy levels of the spectrum of the time-dependent Hamiltonian.
This effect can be explained by the fact that the system is initialized in the instantaneous ground state and has to end up in an excited state after the adiabatic annealing process.
If we only consider the subspace of side-condition-satisfying eigenstates (colored lines in bottom row of Fig.~\ref{fig2}) we are able to see that the system follows the instantaneous conditioned ground state without any level crossings inside this subspace.

In the case that the non-degenerate conditioned ground state is one of the degenerate ground states of the problem without side condition, the unique solution is reached.
For the case that the conditioned ground state $\cgs$ is degenerate, these degenerated states are populated with a bias, influenced by a combination of the choice of the parity layout, the chosen exchange terms and the strength of the parity constraints.
Note that a similar effect was used to program superpositions in unconditioned models using the strength of the parity constraints $C_l$~\cite{siebererlechner2018}.
For more details on degenerate conditioned ground states we refer to Appendix~\ref{appendix_degeneracy}.

Next, we discuss multiple side conditions with different values
$c_j = \sum^{[\mathcal{C}_j]}_i \tilde{\sigma}^{(i)}_z$
for $j=1,... , M$.
If the different side conditions are disjoint in the parity encoding, this method can be extended in a straightforward manner. For details we refer to Appendix~\ref{appendix_multilDisjiontDynamics}. By using ancilla qubits, a parity layout can be constructed, where the qubits included in all given disjoint side conditions are arranged next to each other. In general, one has to pay the price of more qubits and more parity constraints (i.e., physical couplings). 
We give an example for one side condition for which the introduction of an ancilla qubit is necessary to arrange all conditioned qubits next to each other in Appendix~\ref{appendix_parityTransormation}.
The same can be done for multiple side conditions as long as they do not overlap in the parity encoding. We also give an example where two side conditions can be encoded without any ancilla qubits and additional parity constraints.

If two side conditions overlap in one or more parity qubits, this method with the given exchange Hamiltonian~\eqref{eq:Hexchange} cannot be applied directly. However, as long as the complexity of overlapping side conditions is not to high, utilizing Hamiltonians with exchange terms including more than two qubits could be the solution, but this point certainly needs further research.
We provide an example in Appendix \ref{appendix_overlaps}. 

\phantomsection\hypertarget{sec:digital}{}
\section{Digital Quantum Computing}
\label{sec:digital}

In digital quantum computing, an initial state is transferred to a final state via unitary operations. In the quantum approximate optimization algorithm (QAOA)~\cite{farhi2014quantum} these unitaries are constructed from the Hamiltonian encoding the optimization problem and a driver Hamiltonian, usually given by single body $\sigma_x$ terms.
In our approach the unitaries are defined by their corresponding Hamiltonians from Eq.~\eqref{eq:Ht} which also includes exchange terms [see Eq.~\eqref{eq:Hexchange}] as driver terms. For example, the local field term of the problem Hamiltonian given in Eq.~\eqref{eq:localH} corresponds to the unitary
\begin{equation}
	U_{\textrm{final}}(\alpha) = \exp(- i \alpha H_{\textrm{final}}) \text{.}
	\label{eq:QAOAex}
\end{equation}
Each unitary contains a parameter (here $\alpha$) which is variationally updated in the QAOA. 
To simplify the implementation of the unitary corresponding to $H_\textrm{exch.}$ and improve the flexibility of this approach, one can partition the exchange terms such that parts that share qubits are applied sequentially, e.g., 
\begin{eqnarray}
\label{qaoa_mixer_partitioned}
    U_{\textrm{exch.}}(\delta, \epsilon) = &\exp(- i \delta (\tilde{\sigma}_{+}^{(j)} \tilde{\sigma}_{-}^{(k)}  +  \tilde{\sigma}_{-}^{(j)} \tilde{\sigma}_{+}^{(k)}  ))\\\nonumber & \quad \times \exp(- i \epsilon (\tilde{\sigma}_{+}^{(k)} \tilde{\sigma}_{-}^{(l)}  +  \tilde{\sigma}_{-}^{(k)} \tilde{\sigma}_{+}^{(l)}))
\end{eqnarray}
for a side condition including qubits $j$, $k$ and $l$. Mixing operators similar to $U_\textrm{exch.}$ have also been investigated in Refs.~\cite{hadfield2019QAOA, cook2020quantum, qaoa_wang}, although in the context of purely linear side conditions on the logical qubits. 
In contrast to that, this method is applicable to more general side conditions. 

For the simulations presented here we use the partitioned unitary with a single parameter for the different parts (i.e., $\delta = \epsilon$ in Eq.~\eqref{qaoa_mixer_partitioned}, details on the parametrization can be found in Appendix~\ref{appendix_partitioning}).
The final state for a QAOA sequence of length $p$ is then given by
\begin{equation}\label{eq:QAOA_sequence}
\ket{\psi} = \prod_{j=1}^p U_{\textrm{exch.}}(\delta_j) U_{\textrm{x}}(\gamma_j) U_{\textrm{C}}(\beta_j) U_{\textrm{final}}(\alpha_j) \ket{\psi_0}.
\end{equation}
The initial state $\ket{\psi_0}$ has to be prepared in the side-condition-satisfying subspace. This can be achieved by preparing the conditioned qubits in $[\mathcal{C}]$ in a computational basis state that fulfills the side conditions, while the unconditioned qubits in $[\mathcal{U}]$ are prepared in  the $\tilde\sigma_x$-eigenstate $\ket{+}$. Alternatively, one can prepare the conditioned qubits in a Dicke state $\ket{D^n_k}$, where $k$ excitations are symmetrically distributed over all $n$ conditioned qubits. The preparation of such states requires $\mathcal{O}(kn)$ gates~\cite{Baertschi_2019}. To remove any potential bias by the choice of the initial state\footnote{The Hamming distance between a specific computational basis state and the solution of a side condition on $n$ qubits is between $0$ and $n-1$. Thus, starting with a computational basis state can either render the exchange term irrelevant or in the worst case require a sequence-length of $p\geq n-1$ to transfer the excitation to the right place.}, 
the conditioned qubits are initialized in a Dicke state with $k$ chosen to fulfill the corresponding side condition, e.g.,
\begin{equation}
    \ket{\psi_0} = \underbrace{\ket{D_1^3}}_{[\mathcal{C}]} \otimes \underbrace{\ket{+}^{\otimes K-3}}_{[\mathcal{U}]},
\end{equation}
where $K$ is the number of physical qubits. For this example, fulfilling a side condition with target ${\langle c\rangle = 2k-n=-1}$, the Dicke state would be 
\begin{equation*}
    \ket{D_1^3} = \frac{1}{\sqrt{3} }(
      \ket{-1, -1, 1}
    + \ket{-1, 1, -1}
    + \ket{1, -1, -1}).
\end{equation*}
\begin{figure}[t]
\centering
\includegraphics[width=\columnwidth]{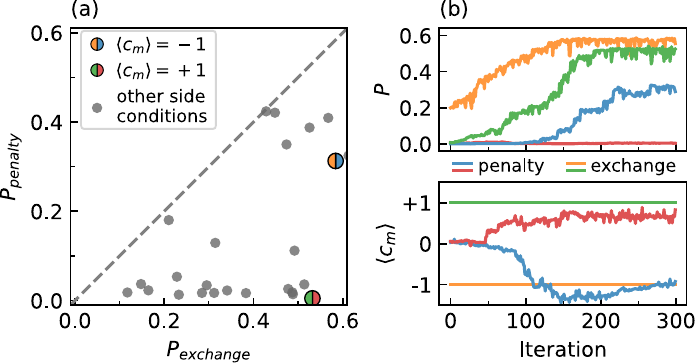}
\caption{Comparison of the QAOA performance using exchange terms and energy penalties to treat side conditions. \textbf{(a)} Scatter plot of the highest reached probabilities to find the lowest-energy state fulfilling the side condition for a given selection of side conditions (see main text) for the \textit{exchange} protocol versus the \textit{penalty} protocol. A point $(P_\textrm{exchange}, P_\textrm{penalty})$ represents the maximal probabilities reached for a specific target-state (given by the side condition and the target value of $\langle c\rangle$) for both protocols, respectively. The side condition $c_m={\sigma_{z}^{(1)}\sigma_{z}^{(2)}+\sigma_{z}^{(2)}\sigma_{z}^{(3)}+\sigma_{z}^{(2)}\sigma_{z}^{(6)}}$ is highlighted for $\langle c_m\rangle=-1$ (\textit{exchange}/\textit{penalty} in orange/blue) and $\langle c_m\rangle=+1$ (green/red).
The panels in \textbf{(b)} show the probability of measuring the target state and the expectation value $\braket{c_m} = \langle\tilde\sigma^{(1,2)}_{z}+\tilde\sigma^{(2,3)}_{z}+\tilde\sigma^{(2,6)}_{z}\rangle$ for the variational state during the iterations of the parameter optimization, for the highlighted cases in panel (a). Note that in the \textit{exchange} protocol $\langle c_m \rangle$ is constant by definition, while the \textit{penalty} protocol has an additional constraint to optimize.
}
\label{fig:QAOA}
\end{figure}
Figure~\ref{fig:QAOA} shows a comparison of this approach (denoted \textit{exchange}) with the fully parallelizable QAOA in the
LHZ architecture~\cite{lechner2015quantum} introduced in~\cite{QAOAWolfgang} (denoted \textit{penalty}), where each side condition is treated by adding an energy penalty to the cost function. We consider the optimization problem in Fig.~\ref{fig1}{\color{quantumviolet}(b)} with the same local fields as in Eq.~\eqref{loc_fields_main}, and take into account all possible side conditions in either a row (e.g., ${\sigma_{z}^{(1)}\sigma_{z}^{(2)}+\sigma_{z}^{(2)}\sigma_{z}^{(3)}+\sigma_{z}^{(2)}\sigma_{z}^{(6)}=c}$) or a column (e.g., ${\sigma_{z}^{(2)}\sigma_{z}^{(3)}+\sigma_{z}^{(3)}\sigma_{z}^{(4)}+\sigma_{z}^{(3)}\sigma_{z}^{(5)}=c}$) and a few additional side conditions which include 4 or 5 qubits (see Appendix~\ref{appendix_qaoa} for a complete list). Here, we explicitly show the performance for different geometries of side conditions for the QAOA because they directly influence the applied gate sequence and the performance.
For a fair comparison only target values of $\langle c \rangle$ which do not uniquely define the state of the conditioned qubits are chosen as side conditions. Otherwise, the exchange term would have no effect and the problem size for the \textit{exchange} protocol would effectively be reduced to just the unconditioned qubits.
The variational state for the \textit{penalty} protocol is given by
\begin{equation}
\label{eq:QAOA_sequence_unconstrained}
\ket{\psi} = \prod_{j=1}^p U_{\textrm{x}}(\gamma_j) U_{\textrm{C}}(\beta_j) U_{\textrm{final}}(\alpha_j) \ket{+}^{\otimes K},
\end{equation}
with the driver unitary
$U_{\textrm{x}}(\gamma_j) = \exp(- i \gamma_j H_{\textrm{x}})$
treating all qubits as being in $[\mathcal{U}]$, i.e., we apply the spin flip driver to all physical qubits.
The cost function is given by ${E = \braket{\psi|H_C+H_\textrm{final}+H_\textrm{penalty}|\psi}}$, with an energy penalty of the form
\begin{equation}
\label{energy_penalty_qaoa}
    H_\textrm{penalty} = C_\textrm{penalty}\left(\sum_i^{[\mathcal{C}]}\tilde{\sigma}_{z}^{(i)} - c_\textrm{target}\right)^2
\end{equation}
for the \textit{penalty} protocol, where $c_\textrm{target}$ denotes the value to which the side condition is restricted to. 
The values of $C_l$ in $H_C$ and
$C_\textrm{penalty}$ in Eq.~\eqref{energy_penalty_qaoa} are set to 4 and we use a QAOA sequence of length $p=2$. The probabilities shown in Fig.~\ref{fig:QAOA} are defined as ${P = |\langle \psi|\psi_\textrm{target}\rangle|^2}$, where $\ket{\psi_\textrm{target}}$ denotes the lowest-energy state fulfilling the side condition.
Details on the parameter optimization and a comparison of different ways to partition $U_\textrm{exch.}$ are available in Appendix~\ref{appendix_qaoa}. 

We observe that both the exchange- and penalty-protocol perform comparably well if the ground state of the unconditioned problem also fulfills the side condition (which is the case for the points in Fig.~\ref{fig:QAOA}{\color{quantumviolet}(a)} with ${P_\textrm{penalty}\geq 0.3}$). For the other cases the \textit{exchange} protocol clearly outperforms the \textit{penalty} protocol. 
In some cases (e.g., the red line in Fig.~\ref{fig:QAOA}{\color{quantumviolet}(b)}) the \textit{penalty} protocol does not converge to the true solution, even though the side condition is approximately fulfilled. This might be remedied by directly including the energy penalty into the compilation process for the physical Hamiltonian (and thus also the QAOA unitaries), which in turn would require more hardware resources. In-depth benchmarking of the QAOA performance and its hardware requirements in these different approaches will be the topic of future research.

\section{Conclusion}
\label{sec:conclusion}
We presented a method to implement and solve optimization problems with additional side conditions in the parity architecture~\cite{lechner2015quantum, compilerpaper}.
The standard approach to implement such side conditions is to embed them as quadratic energy penalties.
Recent papers like Ref.~\cite{a12040077} or Refs.~\cite{ItayHenPRApplied2016, ItayHenPRA2016} suggest
alternative methods to embed pure sum constraints, i.e., ${\sum_{i}\sigma^{(i)}_z=c}$, in a potentially more efficient manner.
Beyond that, this approach enables the embedding of general equality conditions of the form of sums over arbitrary $k$-body products of spin variables, leading to an alternative to the penalty method.
The method given by Ref.~\cite{a12040077} is applicable only for quantum annealing devices based on the Chimera graph.
In Refs.~\cite{ItayHenPRApplied2016, ItayHenPRA2016}, the alternatives to penalty terms are more general, but their experimental feasibility is uncertain. 
Other recent works~\cite{hadfield2019QAOA, cook2020quantum, qaoa_wang} give an alternative to the penalty method for the quantum alternating operator ansatz, using individual mixing Hamiltonians for each class of optimization problems.
Compared to that, in our approach the mixing operators and driver Hamiltonians result from the compilation of a valid parity layout, which means that no mixing terms must be found and the implementation is already given by the encoding step.

We represent side conditions as driver terms in the form of an exchange Hamiltonian that keeps the system subspace that satisfies the side conditions. Our approach can be implemented in the current quantum simulators and quantum computers based on two-dimensional arrays, including Rydberg atoms in optical lattices~\cite{AndersonRydbergLattice,saffman2010quantum, henriet2020quantum,Ebadi_2022,Graham_2022,Bluvstein_2022,Clemens_PhysRevLett.128.120503}, ion surface chips~\cite{britton_engineered_2D_2012,CiracZoller2DarraysIonTraps,Kumph_2011,Mielenz_2016}, superconducting quantum computers~\cite{arute2019quantum,Foxen_2017, gong_2DsuperconductingQubits_2021, menke_2022}, and quantum dots~\cite{quantumdot2D,QD_2017,QD_Veldhorst,QD_Ruoyu},
to name a few.

We compare the encoding of optimization problems in digital and analog applications and find a considerable improvement compared to standard encoding based on energy penalties. Note that in previous architectures, it has been shown that practically implementing such driver terms remains challenging~\cite{choi_minor-embedding_2008, VinciPRA2015}. 

An interesting future outlook is to study the performance of encoded side conditions in the parity architecture in presence of noise or in finite temperature environments.

\section*{Acknowledgements} The numerical simulations for quantum annealing and QAOA throughout this manuscript were performed using the QuTiP package~\cite{Qutip2}.
Work at the University of Innsbruck is supported by the European Union program Horizon 2020 under Grants Agreement No.~817482 (PASQuanS), and by the Austrian Science Fund (FWF) through a START grant under Project No. Y1067-N27 and the SFB BeyondC Project No. F7108-N38, the Hauser-Raspe foundation. This material is based upon work supported by the Defense Advanced Research Projects Agency (DARPA) under Contract No. HR001120C0068. Any opinions, findings and conclusions or recommendations expressed in this material are those of the author(s) and do not necessarily reflect the views of DARPA.


\bibliographystyle{quantum_abrv}


\setcounter{secnumdepth}{2}
\onecolumngrid
\newpage
\twocolumngrid
\appendix
\section*{Appendices}
\addcontentsline{toc}{section}{Appendices}

\section{Parity Transformation\label{appendix_parityTransormation}}
Here, we demonstrate the parity transformation of a given logical problem with side conditions into a suitable parity encoding for some simple examples. Aside from the example given in the main text we consider examples where additional physical qubits are necessary and examples with multiple disjoint side conditions.\\

\paragraph{Transformation of the main example --} In the main text we consider the example of an optimization problem, which can presented by the Hamiltonian
\begin{align}
\label{eq:logical2bodyH}
\nonumber
    H &= \sum_{\langle i,j \rangle} J_{ij} \sigma_z^{(i)}\sigma_{z}^{(j)}\\ \nonumber
    &= J_{12}\sigma_z^{(1)}\sigma_{z}^{(2)} + J_{16} \sigma_z^{(1)}\sigma_{z}^{(6)} \\ \nonumber
    &+ J_{14} \sigma_z^{(1)}\sigma_{z}^{(4)} + J_{26} \sigma_z^{(2)}\sigma_{z}^{(6)} \\ \nonumber
    &+ J_{23} \sigma_z^{(2)}\sigma_{z}^{(3)} + J_{34} \sigma_z^{(3)}\sigma_{z}^{(4)} \\ \nonumber
    &+ J_{35} \sigma_z^{(3)}\sigma_{z}^{(5)} + J_{45} \sigma_z^{(4)}\sigma_{z}^{(5)} \\ 
    &+ J_{56} \sigma_z^{(5)}\sigma_{z}^{(6)}
\end{align}
with the side condition
\begin{equation}
    c =  \sigma_z^{(1)}\sigma_{z}^{(2)} +  \sigma_z^{(2)}\sigma_{z}^{(3)} + \sigma_z^{(2)}\sigma_{z}^{(6)}~.
\end{equation}
The $\sigma_z^{(i)}$ are logical qubits and the $J_{ij}$ are 2-body interaction strenghts.
Each of the products of logical qubits is transformed to one physical qubit
$\sigma_z^{(i)}\sigma_{z}^{(j)} \mapsto \tilde{\sigma}_{z}^{(i,j)}$, such that we have 9 physical qubits in the parity encoding. Each of the interactions maps to a local field $J_{\alpha_{ij}}$ in the parity encoding. The products of logical spins in the side condition are given by the three physical qubits $\tilde{\sigma}_{z}^{(1,2)}$, $\tilde{\sigma}_{z}^{(2,3)}$ and $\tilde{\sigma}_{z}^{(2,6)}$ and the whole side condition in the parity encoding is given by the sum over these physical qubits, i.e.,
\begin{equation}
    c = \tilde{\sigma}_{z}^{(1,2)} + \tilde{\sigma}_{z}^{(2,3)} + \tilde{\sigma}_{z}^{(2,6)}~.
\end{equation}
All of the conditioned qubits are present as interaction terms in the logical graph and therefore it is sufficient to find a parity layout for the problem, where all conditioned qubits are connected via neighbors.
To find the parity constraints we consider first all closed cycles in the logical graph, which are 
\begin{align}
\label{closedcycles1}
    (&\tilde{\sigma}_{z}^{(1,2)}, \tilde{\sigma}_{z}^{(2,3)}, \tilde{\sigma}_{z}^{(3,4)}, \tilde{\sigma}_{z}^{(1,4)}),\\
    \label{closedcycles2}
(&\tilde{\sigma}_{z}^{(1,2)}, \tilde{\sigma}_{z}^{(2,6)}, \tilde{\sigma}_{z}^{(1,6)}),\\
\label{closedcycles3}
(&\tilde{\sigma}_{z}^{(1,4)}, \tilde{\sigma}_{z}^{(1,6)}, \tilde{\sigma}_{z}^{(4,5)}, \tilde{\sigma}_{z}^{(5,6)}),\\
\label{closedcycles4}
(&\tilde{\sigma}_{z}^{(3,4)}, \tilde{\sigma}_{z}^{(4,5)}, \tilde{\sigma}_{z}^{(3,5)}),\\
\label{closedcycles5}
(&\tilde{\sigma}_{z}^{(2,3)}, \tilde{\sigma}_{z}^{(3,5)}, \tilde{\sigma}_{z}^{(5,6)}, \tilde{\sigma}_{z}^{(2,6)}).
\end{align}
The logical problem and side condition in this example have only 2-body terms, which gives us a global spin flip degeneracy ($D=1$). We have $N=6$ logical qubits and $K=9$ parity qubits. This is why we know that we need $K-N+D=9-6+1=4$ parity constraints for a valid parity layout.
If we construct the generator matrix $\mathbf{G}$ and the parity check matrix $\mathbf{P}$ with only 3-body and 4-body rows as explained in Ref.~\cite{compilerpaper}, we are able to find all combinations of four rows of the parity check matrix, which gives us all the possible ways to find a valid parity layout.
From the first four closed cycles [Eqs.~\eqref{closedcycles1}-\eqref{closedcycles4}] listed above we can construct the whole physical lattice, where each of these closed cycles is represented by a parity constraint, which consist either of three or four physical qubits. Furthermore, the choice leads to a physical lattice, in which the conditioned physical qubits are connected via neighbors.
The Hamiltonian in the parity encoding is written as
\begin{align}
\nonumber
    H &= \sum_{k_i} J_{k_i} \tilde{\sigma}_{z}^{(k_i)} \\ \nonumber
    &~~~+ \sum_{\{ijn(l)\} \in P} C_{ijn(l)} \tilde{\sigma}_{z}^{(k_i)}\tilde{\sigma}_{z}^{(k_j)}\tilde{\sigma}_{z}^{(k_n)}(\tilde{\sigma}_{z}^{(k_l)})\\ \nonumber 
    &= J_{12} \tilde{\sigma}_{z}^{(1,2)} + J_{23} \tilde{\sigma}_{z}^{(2,3)} + 
    J_{26} \tilde{\sigma}_{z}^{(2,6)}\\ 
    &~~~+ J_{14} \tilde{\sigma}_{z}^{(1,4)} + J_{34} \tilde{\sigma}_{z}^{(3,4)} + J_{35} \tilde{\sigma}_{z}^{(3,5)} \\  \nonumber
    &~~~+ J_{16} \tilde{\sigma}_{z}^{(1,6)} + J_{45} \tilde{\sigma}_{z}^{(4,5)} + J_{56} \tilde{\sigma}_{z}^{(5,6)}\\ 
    \nonumber
    &~~~+ C_{1234} \tilde{\sigma}_{z}^{(1,2)}\tilde{\sigma}_{z}^{(2,3)}\tilde{\sigma}_{z}^{(3,4)}\tilde{\sigma}_{z}^{(1,4)}\\  \nonumber
    &~~~+ C_{126} \tilde{\sigma}_{z}^{(1,2)}\tilde{\sigma}_{z}^{(2,6)}\tilde{\sigma}_{z}^{(1,6)}\\  \nonumber
    &~~~+ C_{1456} \tilde{\sigma}_{z}^{(1,4)}\tilde{\sigma}_{z}^{(1,6)}\tilde{\sigma}_{z}^{(5,6)}\tilde{\sigma}_{z}^{(4,5)} \\  \nonumber
    &~~~+ C_{345} \tilde{\sigma}_{z}^{(3,4)}\tilde{\sigma}_{z}^{(3,5)}\tilde{\sigma}_{z}^{(4,5)}.
\end{align}
An alternative parity layout in which all conditioned qubits are connected via neighbors can be constructed out of the first two~[Eqs.~\eqref{closedcycles1},\eqref{closedcycles2}] and the last two~[Eqs.~\eqref{closedcycles4},\eqref{closedcycles5}] parity constraints. This alternative parity layout is shown in Fig.~\ref{fig13}{\color{quantumviolet}(a)}.\\

Note that the example considered in the main text without the side condition belongs to the quadratic unconstrained binary optimization (QUBO) problems, i.e., it includes only 2-body interactions. 
For an all-to-all connected QUBO problem the parity encoding is equivalent to the so-called LHZ architecture, which was presented in Ref.~\cite{lechner2015quantum}. There the authors showed, that a fully connected QUBO problem of size $N$ requires $\mathcal{O}(N^2)$ physical qubits in this embedding scheme.
In the case of the parity encoding of a QUBO problem with less than all-to-all connectivity the number of needed parity qubits $K$, including eventually necessary ancilla qubits, can never be bigger than the number of parity qubits which result from an all-to-all connected QUBO problem, since each QUBO problem can always be encoded as a sub-graph of the all-to-all connected graph.
This means, that the parity encoding of a QUBO problem requires $\mathcal{O}(N^2)$ qubits in the worst case.\\

\begin{figure}[!t]
\centering
\includegraphics[width=1\columnwidth]{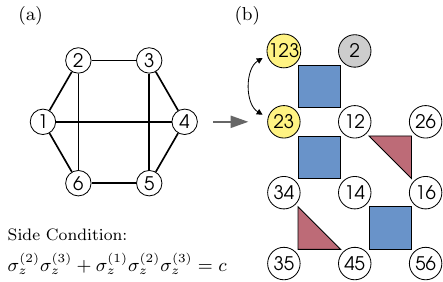}
\caption{
\textbf{(a)} The unconditioned part of the optimization problem is the same as in Fig.~\ref{fig1}{\color{quantumviolet}(a)}. In addition to the optimization problem encoded in the logical graph, the side condition is given separately as ${\sigma_z^{(2)}\sigma_z^{(3)} + \sigma_z^{(1)}\sigma_z^{(2)}\sigma_z^{(3)} =c}$.
\textbf{(b)} In the parity encoding, the encoding of the side condition needs an extra physical qubit for the encoded product ${\sigma_z^{(1)}\sigma_z^{(2)}\sigma_z^{(3)}}$, which is not given in the problem Hamiltonian. Additionally, an ancilla qubit ${\tilde{\sigma}_{z}^{(2)}}$~(grey) together with the additional 4-body parity constraint ${\tilde{\sigma}_{z}^{(2)}\tilde{\sigma}_{z}^{(1,2,3)}\tilde{\sigma}_{z}^{(2,3)}\tilde{\sigma}_{z}^{(1,2)}}$ (blue square) are needed to find a parity layout, which includes the side condition as adjacent qubits~(yellow).}
\label{fig4}
\end{figure}

\paragraph{Transformation for a side condition with additional product terms --\label{appendix_paragraph_scAddTerm}}
If the side condition includes products of logical spins (including single spins), which are not present in the logical Hamiltonian, one has to add more physical qubits and possible ancilla qubits and therefore one needs additional parity constraints, i.e., more physical couplings. Nevertheless, with the help of ancilla qubits it is possible to find a parity encoding for most optimization problems in which the conditioned physical qubits are adjacent. To give an idea on how to do this, an example is presented in Fig.~\ref{fig4}. The examples given here are simple, but the method can be extended to more complicated cases.\\

\paragraph{Transformation of side conditions with the need of ancilla qubits --}
As a first example we take the same Hamiltonian [see Eq.~\eqref{eq:logical2bodyH}] but with the side condition
\begin{equation}
c= \sigma_{z}^{(1)}\sigma_{z}^{(2)}\sigma_{z}^{(3)} + \sigma_{z}^{(2)}\sigma_{z}^{(3)}~.
\end{equation}
For this example we have to add the ancilla qubit $\tilde{\sigma}_z^{(2)}$ to ensure a valid encoding. We get the same chip as in Fig.~\ref{fig1} with the two additional physical qubits $\tilde{\sigma}_z^{(2)}$ and $\tilde{\sigma}_z^{(1,2,3)}$, both with zero local field, and the additional parity constraint consisting of the four physical qubits 
$\tilde{\sigma}_z^{(2,3)}$, $\tilde{\sigma}_z^{(1,2)}$, $\tilde{\sigma}_z^{(2)}$ and $\tilde{\sigma}_z^{(1,2,3)}$. Figure \ref{fig4} shows the logical problem and the corresponding parity encoding.\\

\begin{figure}[!t]
\centering
\includegraphics[width=1\columnwidth]{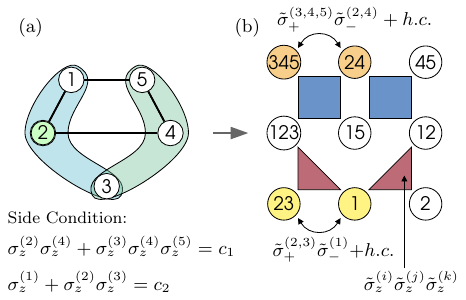}
\caption{%
\textbf{(a)} The unconditioned part of the optimization problem is encoded in a hypergraph (top).
In addition, two side conditions on the spins are given (bottom). They do overlap in the logical qubits $\sigma^{(2)}_{z}$ and $\sigma^{(3)}_{z}$.
\textbf{(b)} In the parity encoding, the problem is translated to a Hamiltonian that encodes the unconditioned as well as the conditioned part. The qubits of the conditioned part are connected via adjacent conditioned qubits (yellow and orange).
In the parity encoding the side conditions are pure sums over single parity qubits and they do not overlap.
}
\label{fig5}
\end{figure}

\paragraph{Two disjoint side conditions --\label{appendix_paragraph_disjointSC}}
A simple example for the case of two disjoint (in the parity encoding) side conditions could be the following Hamiltonian, which has single, 2-body and 3-body terms
\begin{align}
\label{eq:complexH}
\nonumber
    H &= J_{2} \sigma_z^{(2)} + J_{12}\sigma_z^{(1)}\sigma_{z}^{(2)} \\ \nonumber
    &+ J_{15}\sigma_z^{(1)}\sigma_{z}^{(5)} + J_{24}\sigma_z^{(2)}\sigma_{z}^{(4)} \\ \nonumber
    &+ J_{45}\sigma_z^{(4)}\sigma_{z}^{(5)}
    + J_{123}\sigma_z^{(1)}\sigma_{z}^{(2)}\sigma_{z}^{(3)}\\ 
    &+ J_{345}\sigma_z^{(3)}\sigma_{z}^{(4)}\sigma_{z}^{(5)}
\end{align}
with the two side conditions
\begin{align}
    c_1 &= \sigma_z^{(2)}\sigma_z^{(4)} + \sigma_z^{(3)}\sigma_z^{(4)}\sigma_z^{(5)},\\
    c_2 &= \sigma_z^{(1)} + \sigma_z^{(2)}\sigma_z^{(3)}
\end{align}
in the logical encoding.
The side conditions in the parity encoding are again only sums of single physical qubits
\begin{align}
    c_1 &= \tilde{\sigma}_{z}^{(2,4)} + \tilde{\sigma}_{z}^{(3,4,5)},\\
    c_2 &= \tilde{\sigma}_{z}^{(1)} + \tilde{\sigma}_{z}^{(2,3)}.
\end{align}
While the logical side conditions do overlap in the single qubits $\sigma_z^{(2)}$ and $\sigma_z^{(3)}$,
in the parity encoding the side conditions are disjoint, because of the fact that each product of logical qubits transforms to a single parity qubit.
In general, as long as the terms of multiple side conditions consist of different products of logical spins, the side conditions in the parity encoding are disjoint.
The Hamiltonian given in Eq.~\eqref{eq:complexH} is presented in Fig.~\ref{fig5}{\color{quantumviolet}(a)} as a logical graph.
In the parity encoding, the Hamiltonian and the side conditions are both given in the physical graph. Each product of logical qubits in the side conditions which does not exist in the logical problem Hamiltonian needs an extra physical qubit. All physical qubits are composed with parity constraints to a parity layout. The parity constraints again ensure  a clear correspondence between logical and physical states.
In this example a transformation even without the use of ancilla qubits is possible.\\

\paragraph{General procedure with side conditions --}
For any logical Hamiltonian with side conditions we can use the parity transformation as it is described in Ref.~\cite{compilerpaper}. We use the concepts of the generator matrix $\mathbf{G}$ and parity check matrix $\mathbf{P}$ to describe the parity transformation as introduced in Ref.~\cite{compilerpaper}.
First we set up a new generator matrix $\mathbf{G}$, by adding a column for each product of logical spins which are only given in the side condition. Then we proceed as described in Ref.~\cite{compilerpaper} to find a suitable parity check matrix $\mathbf{P}$ and from there a parity layout. In the last two steps it could be necessary to introduce ancilla qubits.
The only difference in finding a valid parity check matrix $\mathbf{P}$ compared to the case without side conditions is that here one has to select rows of $\mathbf{P}$ with the extra condition that the qubits included in the side condition can be arranged adjacent in the parity layout. If this is not possible, we have to add ancilla qubits such that this is possible.
The parity transformation can also be applied in the presence of side conditions
for optimization problems with Hamiltonians represented by hypergraphs (i.e., interactions including more than two qubits).
In the parity encoding each hyperedge of degree $k$ is encoded by one physical qubit $\tilde{\sigma}_z^{(i_1, i_2, ..., i_k)} = \sigma_z^{(i_1)}\sigma_z^{(i_2)}...\sigma_z^{(i_k)}$ and with a sufficient number of suitable ancilla qubits one can find a parity layout, in which every side condition is presented by parity qubits, which are arranged next to each other.
Note, that the minimal number of ancilla qubits necessary for compiling with side conditions is still an open question and part of ongoing research, even for the case that the underlying optimization problem is a QUBO problem.
In some cases, such as a side condition having no or exactly one solution, preprocessing can be used to solve the problem more efficiently.
In the following we discuss some special cases of interrelation between the parity constraints and side conditions including further demonstrations of the parity transformation in the presence of side conditions.

\paragraph{Decoding --}
For the process of decoding a parity encoded state back to a logical qubit state we refer to the corresponding section in Ref.~\cite{compilerpaper}.
In an ideal device without any noise that procedure can be used for both cases, with and without additional hard side conditions.
In the presence of device dependent errors one has to use different decoding strategies.
In that case, some of the parity constraints can be violated at the end of the computation, which can lead to contradictions in the decoding process of the logical state.
For the case without hard side conditions different error correcting/mitigating decoding strategies have already been proposed, e.g., for quantum annealing devices as explained in Refs.~\cite{albashLidarMEvsLHZ, BeliefPropLHZpreskill} and for digital ones using the QAOA as explained recently in Ref.~\cite{weidingerMBengErrorQAOA}.
For the method presented here, which encodes additional hard side conditions for both, analog and digital devices, the decoding strategies have to be adapted.
If the side conditions are not taken into account, these decoding strategies can lead to unsatisfied side conditions in the decoded state.
In the strategy for QAOA presented in Ref.~\cite{weidingerMBengErrorQAOA} one could for example only use spanning trees, which cover all conditioned qubits in the parity encoding. 
For belief propagation, as suggested in Ref.~\cite{BeliefPropLHZpreskill} as a decoding strategy for quantum annealing, one could fix the probabilities of all conditioned qubits to be in their respective state to 1
and never update these probabilities during the iteration of belief propagation.
This should give the reader an idea that small changes in the existing decoding strategies for noisy devices can conserve the side conditions.
The details and the performance of the decoding strategies which include side conditions in the presence of errors 
is a part of future research work.

\subsection{Interrelation between parity constraints and side conditions\label{appendix_InterplayParityConstraintsSideConditions}}
There are four cases in which one can think of interrelations between parity constraints and side conditions.

\begin{figure}[!t]
\centering
\includegraphics[width=1\columnwidth]{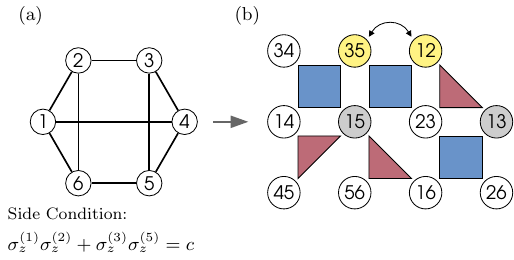}
\caption{\textbf{(a)}
The unconditioned part of the optimization problem is the same as in Fig.~\ref{fig1}. The side condition is given separate as $\sigma_z^{(1)}\sigma_z^{(2)} + \sigma_z^{(3)}\sigma_z^{(5)} =c$.
\textbf{(b)} In the parity encoding, the problem is translated to a Hamiltonian that encodes the unconditioned as well as the conditioned part. The side condition is implemented as exchange term among adjacent conditioned parity qubits~(yellow). We need two ancilla qubits~(grey), and therefore two additionally parity constraints, to lay out the side conditioned parity qubits as neighbors, although no additional parity qubits for the product terms in the side conditions itself are needed.
}
\label{fig6}
\end{figure}

\paragraph{Arranging conditioned qubits --} It can be the case that there is no parity layout for the physical qubits representing the given interaction terms and side conditions where the side conditioned physical qubits are connected via adjacent qubits. In this case one has to add ancilla qubits to obtain a layout in which the side conditioned qubits are adjacent and therefore additional parity constraints are needed. In Fig.~\ref{fig6} we give an example, where we have to introduce at least the two ancilla qubits $\tilde{\sigma}_z^{(1,5)}$ and $\tilde{\sigma}_z^{(1,3)}$, to arrange $\tilde{\sigma}_z^{(1,2)}$ next to $\tilde{\sigma}_z^{(3,5)}$. Moreover, the two ancilla qubits require two additional parity constraints for a valid layout.
\paragraph{Conditioned qubits that do not appear in the problem Hamiltonian --}
As it is explained above in Appendix~\ref{appendix_paragraph_scAddTerm}, in general new spin products in the side condition lead to additional parity qubits and therefore to additional parity constraints (see Fig.~\ref{fig4}).
But in the case that a side condition is a single product term, it can be subsumed into the parity constraints and it is not necessary to introduce additional qubits or to use more parity constraints. For more details we refer to Ref.~\cite{compilerpaper}.

\paragraph{Violated parity constraints --}
The next cases we want to discuss are potential conflicts between parity constraints and side conditions.
The first case is that a side condition can only be fulfilled if one or more parity constraints are violated.
The second case is that multiple disjoint side conditions with neighboring parity constraints can only be fulfilled at the same time if one or more parity constraints are violated.
With the dynamics presented here, we will start and end up in a system state, which fulfills the side conditions but has at least one violated parity constraint. But both cases are not real conflicts, because in these cases an invalid parity constraint means that the side conditions are not solvable.
With classical preprocessing one can exclude these cases before the compilation process.
Every side condition over binary variables as given in Eq.~\eqref{eq:sumconst} can be rewritten in a conjunctive normal form (CNF) and checking if a CNF can be satisfied is a boolean satisfiability problem (SAT), which is known to be NP-complete.
Since checking if a side condition is satisfiable is NP-complete, classical preprocessing is not an option for every side condition. 
We want to note, that the method presented here is mainly intended for optimization problems with additional side conditions, where the focus is not on the satisfiability of the side condition itself.
In a large amount of these problems the satisfiability of the additional side conditions is presumed or can be easily checked, which for example is the case for the class of scheduling problems~\cite{venturelli_quantum_2016_chop_shop, ikeda_application_2019}.

To give an example for each of the two cases we take the logical Hamiltonian in Fig.~\ref{fig1} for both examples. For the first example we think of the side condition
\begin{equation}
    \label{invalideSC1}
    -2 = \langle \sigma_z^{(1)}\sigma_z^{(2)} + \sigma_z^{(2)}\sigma_z^{(3)}
    + \sigma_z^{(3)}\sigma_z^{(4)} + \sigma_z^{(1)}\sigma_z^{(4)} \rangle
\end{equation}
or in the parity encoding
\begin{equation}
    \label{invalideSC1PC}
    -2 = \langle\tilde{\sigma}_z^{(1,2)}  + \tilde{\sigma}_z^{(2, 3)} + \tilde{\sigma}_z^{(3, 4)}+ \tilde{\sigma}_z^{(1,4)}\rangle ~.
\end{equation}
This side condition is only fulfilled if three of the physical qubits have eigenvalue -1, which in any case leads to a violated parity constraint.
But on the other hand, we note at this point that if a parity layout exists, in which the conditioned parity qubits build exact one existing 3- or 4-body constraint, the corresponding parity constraint can be dropped. This means the side condition leads to fewer physical couplings.

\begin{figure}[!t]
\centering
\includegraphics[width=1\columnwidth]{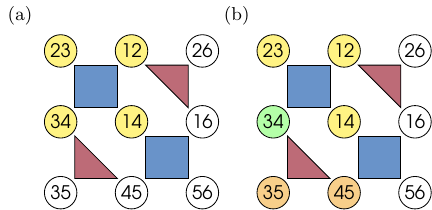}
\caption{
Parity encoding of the Hamiltonian given in Eq.~\eqref{eq:logical2bodyH}
\textbf{(a)} with side condition $-2 = \langle\sigma_z^{(1)}\sigma_z^{(2)} + \sigma_z^{(2)}\sigma_z^{(3)}
    + \sigma_z^{(3)}\sigma_z^{(4)} + \sigma_z^{(1)}\sigma_z^{(4)} \rangle$ (yellow nodes), which has no solution and can only be fulfilled if the corresponding parity constraint (upper left blue square) is violated.
\textbf{(b)} with the two side conditions $-2 = \langle\sigma_z^{(3)}\sigma_z^{(5)} + \sigma_z^{(4)}\sigma_z^{(5)}\rangle$ (orange nodes) and
$-3 = \langle\sigma_z^{(1)}\sigma_z^{(2)} + \sigma_z^{(2)}\sigma_z^{(3)}
    + \sigma_z^{(1)}\sigma_z^{(4)}\rangle$ (yellow nodes). The both side conditions can not be fulfilled at the same time. In the parity encoding both side conditions are fulfilled only if one of the two adjacent parity constraints is violated. The qubit $\tilde{\sigma}_z^{(3,4)}$ (green node), which is neither in no one of the side conditions has a conflict. It can be adjusted, such that only one of two parity constraints is not violated.
}
\label{fig7}
\end{figure}

For the second example we can think of the two side conditions
\begin{align}
    \label{invalideSC2}
    -2 &= \langle\sigma_z^{(3)}\sigma_z^{(5)} + \sigma_z^{(4)}\sigma_z^{(5)}\rangle,\\
    -3 &= \langle\sigma_z^{(1)}\sigma_z^{(2)} + \sigma_z^{(2)}\sigma_z^{(3)}
    + \sigma_z^{(1)}\sigma_z^{(4)}\rangle
\end{align}
or in the parity encoding
\begin{align}
    \label{invalideSC2PC}
    -2 &= \langle\tilde{\sigma}_z^{(3,5)} + \tilde{\sigma}_z^{(4,5)}\rangle,\\
    -3 &= \langle\tilde{\sigma}_z^{(1,2)}  + \tilde{\sigma}_z^{(2, 3)}+ \tilde{\sigma}_z^{(1,4)}\rangle.
\end{align}
Both side conditions are only fulfilled if one of the involved parity constraints is violated.
In Fig.~\ref{fig7} we illustrate both cases.

If a side condition is solvable and the annealing process ends up in a parity constraint violating state, the parity constraints strength was chosen too low and has to be increased.

\paragraph{Degenerate groundstates --}
The last case of interrelation, is only present if the conditioned problem has a degenerated ground state manifold. In this case, the probabilities of the degenerated ground states are biased due to the combination of the choice of the exchange Hamiltonian and the given parity constraints and the corresponding parity layout. As shown in Ref.~\cite{siebererlechner2018}, the probabilities of degenerated ground states in unconditioned problems are biased by the values of the parity-constraint strengths. In a similar manner, we can show how the combination of parity layout, exchange Hamiltonian and parity constraints influences the ground state populations for problems with side conditions.
We will explain it with an example in more detail in Appendix~\ref{appendix_degeneracy}.

\subsection{Overlapping side conditions \label{appendix_overlaps}}
If side conditions are not disjoint (see Appendix~\ref{appendix_paragraph_disjointSC}) we call them overlapping side conditions.
As we show in the example given in Fig.~\ref{fig5}, side conditions
only overlap in the parity encoding if they contain the same product of logical qubits.
Therefore, we again consider the Hamiltonian from Eq.~\eqref{eq:complexH}, but with the following two side conditions:
\begin{align}
    c_1 &= \sigma_z^{(2)}\sigma_z^{(4)} + \sigma_z^{(3)}\sigma_z^{(4)}\sigma_z^{(5)} + \sigma_z^{(1)}\sigma_z^{(2)}\sigma_z^{(3)},\\
    c_2 &= \sigma_z^{(1)} + \sigma_z^{(2)}\sigma_z^{(3)} + \sigma_z^{(1)}\sigma_z^{(2)}\sigma_z^{(3)}.
\end{align}
The product $\sigma_z^{(1)}\sigma_z^{(2)}\sigma_z^{(3)}$ is present in both side conditions.
In the parity encoding this translates to the side conditions
\begin{align}
    c_1 &= \tilde{\sigma}_{z}^{(2,4)} + \tilde{\sigma}_{z}^{(3,4,5)} + \tilde{\sigma}_{z}^{(1,2,3)},\\
    c_2 &= \tilde{\sigma}_{z}^{(1)} + \tilde{\sigma}_{z}^{(2,3)} + \tilde{\sigma}_{z}^{(1,2,3)}~,
\end{align}
which overlap in the physical qubit $\tilde{\sigma}_{z}^{(1,2,3)}$.
Because the exchange Hamiltonian~\eqref{eq:Hexchange} conserves the total sum of all spins in all side conditions combined, applying the exchange Hamiltonian as given in Eq.~\eqref{eq:Hexchange} is not enough to stay in the side-condition-satisfying subspace.
A straightforward way to extend the method presented here to the case of side conditions with overlapping physical qubits is to replace the 2-body exchange Hamiltonian given in Eq.~\eqref{eq:Hexchange} by an exchange Hamiltonian with higher body terms.
For the case of an overlap in a single qubit it is enough to use an exchange Hamiltonian with 2- and 3-body terms, for example
\begin{align}
\label{eq:exOverlapH}\nonumber
    H_{\textrm{exch.}} &= \tilde{\sigma}_{+}^{(2,4)} \tilde{\sigma}_{-}^{(3,4,5)}
    + \tilde{\sigma}_{+}^{(1)} \tilde{\sigma}_{-}^{(2,3)}+ \\
    &+ \tilde{\sigma}_{+}^{(1,2,3)}\tilde{\sigma}_{-}^{(2,3)}\tilde{\sigma}_{-}^{(3,4,5)}+ h.c.
\end{align}
Depending on the number of overlapping side conditions and the number of qubits in which they overlap it could be difficult to find a parity layout with only 3- and 4-body parity constraints in which all qubits of one side condition are neighboring, e.g., if we have more than 4 side conditions, which all overlap in the same physical qubit.
In the case that each side condition has no coefficients different from 1 (i.e., all ${g^{(m)}_{\alpha}= 1}$) and that the conditioned qubits of one side condition include all qubits of the other side condition,
the side conditions can be transformed to disjoint side conditions via Gaussian elimination.
\subsection{Side conditions with integer coefficients}
In the case of integer coefficients (i.e., at least one ${g^{(m)}_{\alpha}\neq 1}$) in the side condition terms one can use the presented method again by an exchange Hamiltonian with higher body terms. As an example, we again use the Hamiltonian given in Eq.~\eqref{eq:complexH}, but with the side condition
\begin{equation}
    c = \tilde{\sigma}_{z}^{(1)} + 2\tilde{\sigma}_{z}^{(2,3)} + 3\tilde{\sigma}_{z}^{(1,2,3)}~.
\end{equation}
For this side condition the method presented in this paper would work with an exchange Hamiltonian of the form
\begin{equation}
\label{eq:prefactorHex}
    H_{\textrm{exch.}} = \tilde{\sigma}_{+}^{(1,2,3)}\tilde{\sigma}_{-}^{(2,3)}\tilde{\sigma}_{-}^{(1)}+ h.c.
\end{equation}
For the side condition
\begin{equation}
    c = \tilde{\sigma}_{z}^{(1)} + \tilde{\sigma}_{z}^{(2,3)} + 2\tilde{\sigma}_{z}^{(1,2,3)}~
\end{equation}
one would require an exchange Hamiltonian of the following form:
\begin{align}
    \label{eq:prefactorHex3}
    H_{\textrm{exch.}} &= \tilde{\sigma}_{+}^{(1)} \tilde{\sigma}_{-}^{(2,3)}
    + \tilde{\sigma}_{+}^{(1,2,3)}\tilde{\sigma}_{-}^{(1)}\tilde{\sigma}_{-}^{(2,3)} + h.c.
\end{align}
In the case of a side condition
\begin{equation}
    c = \tilde{\sigma}_{z}^{(1)} + \tilde{\sigma}_{z}^{(2,3)} + 3\tilde{\sigma}_{z}^{(1,2,3)}
\end{equation}
one has either no solution or exactly one, depending on the value of $\langle c\rangle$. In this case it is trivial to see that no exchange Hamiltonian is required and instead we can fix the three qubits in the side condition, such that they fulfill the side condition.

%
%
\subsection{Inequalities \label{inequalities}}
Inequalities can be written as multiple equalities at the cost of extra qubits, as was shown in Ref.~\cite{lucas2014ising}. In the parity encoding, inequalities lead to multiple overlapping equality side conditions where some coefficients take on values ${g^{(m)}_{\alpha}\neq 1}$. Here we give a simple example for an inequality.
The logical optimization problem for this example with $N=4$ logical qubits is given by 
\begin{align}\nonumber
	H_{\textrm{problem}} &= J_{12} \sigma_{z}^{(1)}  \sigma_{z}^{(2)}
	+ J_{13} \sigma_{z}^{(1)}  \sigma_{z}^{(3)}\\
	&+ J_{24} \sigma_{z}^{(2)}  \sigma_{z}^{(4)}
    +J_{34} \sigma_{z}^{(3)}  \sigma_{z}^{(4)}
\end{align}
and the inequality is given by
\begin{equation}
	\sum_{i}^{3} \langle\sigma_{z}^{(i)} \rangle \leq \gamma~, ~\gamma \in \mathbb{Z}.
\end{equation}
First we follow Ref.~\cite{lucas2014ising} to bring this inequality into a system of equalities.
Here, the expectation values of logical qubits are $\langle\sigma_{z}^{(i)} \rangle\in \{\pm1\}$ and the sum $\sum_{i}^{3}\langle \sigma_{z}^{(i)}\rangle$ can take a discrete number $m$ of different values, which depend on $\gamma$.
Here, for $\gamma < -3$ it is $m=0$ and no solution exists, for $-3 \leq \gamma <-1$ it is $m=1$, for $-1 \leq \gamma < 1$ it is $m =2$, for $1 \leq \gamma < 3$ it is $m=3$ and for $3 \leq \gamma $ it is $m =3$, but the side condition will not restrict the space of solutions any longer.

\begin{figure}[!t]
	\centering
	\includegraphics[width=1\columnwidth]{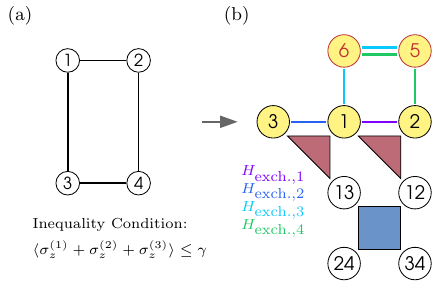}
	\caption{ 
	\textbf{(a)} The unconditioned part of the optimization problem is encoded as a spin model (top). Each node (circle) represents a logical spin, which can either be up or down. The edges (lines) represent interactions between the spins. The solution of the unconditioned part is the ground state of the spin model. In addition, a side condition in the form of an inequality on the spins is given (bottom). \textbf{(b)} In the parity encoding, the side conditions are implemented as exchange terms among parity qubits that are part of the first side condition (filled yellow) and the second side condition (colored red), which together represent the inequality side condition. The exchange terms among physical qubits which are not in the overlapping part can be realized by two body exchange terms, as long as they have the same coefficient in the side condition. For this example we need four different exchange terms (four different colors of lines connecting conditioned qubits), two 2-body and two 3-body exchange terms, given in Eq.~\eqref{eq:Hexfull}.}
	\label{fig8}
\end{figure}

We introduce $m$ ancilla qubits $\sigma_{z}^{(j)}$ with $j=N+1, ..., N+m$, which have to fulfill the equalities
\begin{equation}
	\label{eq:eq1}
	\sum_{i=1}^{m}	\langle\sigma_{z}^{(i+N)}\rangle  - (m - 2) = 0
\end{equation}
and 
\begin{equation}
	\label{eq:eq2}
	\frac{1}{2} \sum_{i=1}^{m} \beta_i \left(1+\langle \sigma_{z}^{(i+N)}\rangle\right) -  \sum_{i = 1}^{3} \langle \sigma_{z}^{(i)} \rangle= 0.
\end{equation}
The values $\beta_i$ are the discrete values such that $\sum_{j}^{3} \langle\sigma_{z}^{(j)}\rangle=\beta_i$ exists.
These two equalities together replace the inequality. The second equality can be written as
\begin{equation}
	\label{eq:eq2var2} \sum_{i=1}^{m} \beta_i \langle\sigma_{z}^{(i+N)}\rangle - 2 \sum_{i = 1}^{3} \langle\sigma_{z}^{(i)}\rangle = -\sum_{i=1}^{m} \beta_i ,
\end{equation}
which we will use later.
Here we show a possible parity encoding with side conditions for the case $\gamma = 0$.
In this case we have $m=2$, $\beta_1 = -3$ and $\beta_2=-1$ and we need two logical ancilla qubits to transform the inequality into the two equalities
\begin{align}
    \nonumber
    \langle\sigma_{z}^{(5)}+\sigma_{z}^{(6)}\rangle &= 0,\\
    2\langle \sigma_{z}^{(1)}+\sigma_{z}^{(2)}+\sigma_{z}^{(3)}
    \rangle + 3\langle\sigma_{z}^{(5)}\rangle + \langle \sigma_{z}^{(6)}\rangle &= 4 .
\end{align}
For the parity encoding we get the four physical qubits $\tilde{\sigma}_{z}^{(1,2)}, \tilde{\sigma}_{z}^{(1,3)}, \tilde{\sigma}_{z}^{(2,4)}$ and $\tilde{\sigma}_{z}^{(3,4)}$ from the optimization problem, the three physical qubits $\tilde{\sigma}_{z}^{(1)}, \tilde{\sigma}_{z}^{(2)}$ and $\tilde{\sigma}_{z}^{(3)}$ from the inequality and two physical qubits $\tilde{\sigma}_{z}^{(5)}$ and $\tilde{\sigma}_{z}^{(6)}$ from the ancilla qubits to write the inequality in form of two equalities. For a valid parity transformation we need no further ancilla qubits in this case.
The physical qubits $\tilde{\sigma}_{z}^{(5)}$ and $\tilde{\sigma}_{z}^{(6)}$ included in equality~\eqref{eq:eq1} are neighboured and all other conditioned qubits are connected via adjacent qubits in the given parity layout.
A valid parity layout for this example is shown in Fig.~\ref{fig8}{\color{quantumviolet}(b)}.

The two side conditions are overlapping and thus we need an exchange Hamiltonian with higher-body terms,
which conserves both side conditions. We have to take into account that the physical qubits in the second side conditions have coefficients ${g^{(m)}_{\alpha}\neq 1}$. Flipping qubit $\tilde{\sigma}_{z}^{(5)}$ changes the sum by a factor of 3, flipping $\tilde{\sigma}_{z}^{(6)}$ gives a change of 1 and all the other ones lead to a change of 2. We are able to find in total six exchange terms among nearest neighbors, thereof two 2-body terms, two 3-body terms and three 4-body terms.
\begin{align}
\label{eq:Hex1}
    H_{\textrm{exch.},1} &= \tilde{\sigma}_{+}^{(1)}\tilde{\sigma}_{-}^{(2)}  + h.c.,\\
    \label{eq:Hex2}
    H_{\textrm{exch.},2} &= \tilde{\sigma}_{+}^{(1)}\tilde{\sigma}_{-}^{(3)}  + h.c.,\\
    \label{eq:Hex3}
    H_{\textrm{exch.},3} &= \tilde{\sigma}_{+}^{(5)}\tilde{\sigma}_{-}^{(6)} \tilde{\sigma}_{-}^{(1)} + h.c.,\\
    \label{eq:Hex4}
    H_{\textrm{exch.},4} &= \tilde{\sigma}_{+}^{(5)}\tilde{\sigma}_{-}^{(6)} \tilde{\sigma}_{-}^{(2)}  + h.c.,\\
    \label{eq:Hex5}
    H_{\textrm{exch.},5} &= \tilde{\sigma}_{+}^{(1)}\tilde{\sigma}_{-}^{(5)} \tilde{\sigma}_{-}^{(6)}\tilde{\sigma}_{+}^{(2)}  + h.c.,\\
    \label{eq:Hex6}
    H_{\textrm{exch.},6} &= \tilde{\sigma}_{+}^{(1)}\tilde{\sigma}_{-}^{(5)} \tilde{\sigma}_{-}^{(6)}\tilde{\sigma}_{+}^{(3)} + h.c.
\end{align}
The terms~\eqref{eq:Hex5} and~\eqref{eq:Hex6} can be dropped, because the first side condition~\eqref{eq:eq1} only allows one of the qubits $\tilde{\sigma}_{z}^{(5)}$, $\tilde{\sigma}_{z}^{(6)}$ to be in the state with eigenvalue 1, which means applying terms~\eqref{eq:Hex5} and~\eqref{eq:Hex6} on states, which fulfill both side conditions will give a zero contribution.
The terms~\eqref{eq:Hex1},~\eqref{eq:Hex2},~\eqref{eq:Hex3} and~\eqref{eq:Hex4} are enough to reach every state, which fulfills both side conditions. All eigenstates, which fulfill both side conditions have one of the four alignments of the subset
$(\langle \tilde{\sigma}_{z}^{(1)}\rangle ,
\langle \tilde{\sigma}_{z}^{(2)}\rangle,
\langle \tilde{\sigma}_{z}^{(3)}\rangle,
\langle \tilde{\sigma}_{z}^{(5)}\rangle,
\langle \tilde{\sigma}_{z}^{(6)}\rangle)$:
\begin{equation}
    (\langle\tilde{\sigma}_{z}^{(1)}\rangle \dots \langle\tilde{\sigma}_{z}^{(6)}\rangle) \in 
    \begin{cases}
        (-1, -1, -1, ~~1, -1)&= \alpha_1 \\
        (~~1, -1, -1, -1, ~~1)&= \alpha_2 \\
        (-1, ~~1, -1, -1, ~~1)&= \alpha_3 \\
        (-1, -1, ~~1, -1, ~~1)&= \alpha_4
    \end{cases}
\end{equation}
Let us call them states of type $\alpha_i$ and write them as $\ket{\phi^{\alpha_i}}$. Then $H_{\textrm{exch.},1}$ maps between $\ket{\phi^{\alpha_2}}$ and $\ket{\phi^{\alpha_3}}$, $H_{\textrm{exch.},2}$ maps between $\ket{\phi^{\alpha_2}}$ and $\ket{\phi^{\alpha_4}}$, $H_{\textrm{exch.},3}$ maps between $\ket{\phi^{\alpha_1}}$ and $\ket{\phi^{\alpha_2}}$ and $H_{\textrm{exch.},4}$ maps between $\ket{\phi^{\alpha_1}}$ and $\ket{\phi^{\alpha_3}}$.
This means that with the exchange Hamiltonian
\begin{align}
    \label{eq:Hexfull}
    H_{\textrm{exch.}} &= H_{\textrm{exch.},1} + H_{\textrm{exch.},2} \\ \nonumber
    &+ H_{\textrm{exch.},3} + H_{\textrm{exch.},4}\\ \nonumber
    &= \tilde{\sigma}_{+}^{(1)}\tilde{\sigma}_{-}^{(2)} + 
    \tilde{\sigma}_{+}^{(1)}\tilde{\sigma}_{-}^{(3)} \\ \nonumber
    &+ \tilde{\sigma}_{+}^{(5)}\tilde{\sigma}_{-}^{(6)} \tilde{\sigma}_{-}^{(1)} +
    \tilde{\sigma}_{+}^{(5)}\tilde{\sigma}_{-}^{(6)} \tilde{\sigma}_{-}^{(2)} + h.c.
\end{align}
and the Hamilton dynamics presented in the main text, we can reach each state that lies in the subspace of side-condition-fulfilling eigenstates. As we never leave this subspace after appropriately choosing the initial state we are able to find the lowest energy state that fulfills the inequality.
The experimental implementation of such higher body exchange terms in quantum annealing is unclear and further research is needed for this point.

\section{Annealing protocol \label{appendix_annealingprotocol}}
\begin{figure}[htb]
	\centering
\includegraphics[width=0.85\columnwidth]{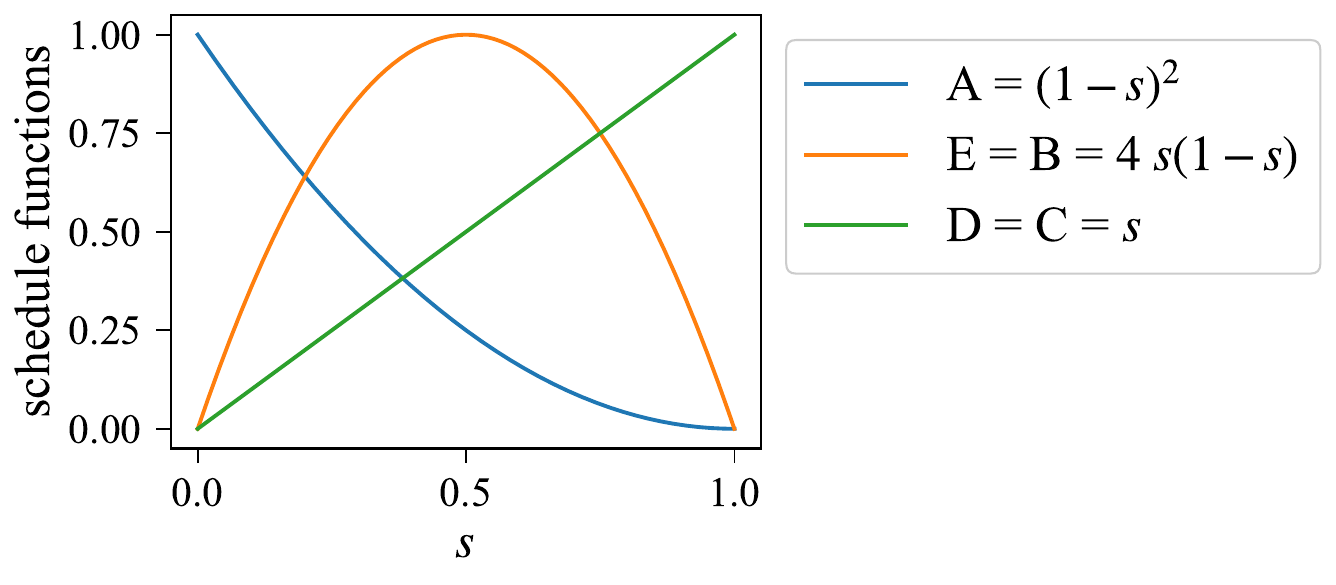}
\caption{%
The annealing schedule used for the numerical quantum annealing examples in this work.
}
\label{fig9}
\end{figure}

The time dependent protocol used in the numerical examples is given by
${A(s) = (1-s)^2}$, ${B(s) = E(s)=\Gamma s(1-s)}$ and ${C(s) = D(s) = s}$, where ${s=t/t_f \in [0, 1]}$ and $t_f$ is the final annealing time of the annealing process, illustrated in Fig.~\ref{fig9}. Therefore, the time-dependent Hamiltonian given in Eq.~\eqref{eq:Ht} explicitly written-out is given by
\begin{eqnarray}
\label{eq:Hs}
\nonumber
	H(s) = & (1-s)^2& ~H_{\textrm{init}}\\
	+ & ~\Gamma s(1-s)~& \left(H_{\textrm{x}} + H_{\textrm{exch.}} \right)\\ \nonumber
	+ & s&  \left(H_{\textrm{final}} + H_{\textrm{C}} \right).
\end{eqnarray}
The factor $\Gamma$ has to be chosen suitably. If $\Gamma$ is too small the energy gap between the conditioned ground state and the next higher conditioned state may close. This means that large annealing times are needed in order to stay in the conditioned ground state during the annealing process.
A large $\Gamma$ introduces a new energy scale and the advantage 
over the method using penalty terms vanishes.
Here we use $\Gamma =4$ for all examples.
The  strength  of the  parity  constraints is set to $C_l = -4$ and the exchange term constant is set to $\gamma_e = 1/2$ for all examples. 

\section{Hamiltonian dynamics for multiple disjoint side conditions
\label{appendix_multilDisjiontDynamics}}
The dynamics presented in the main text can be easily extended to multiple disjoint side conditions. The Hamiltonians $H_{\textrm{final}}$, $H_{\textrm{C}}$ and $H_{\textrm{x}}$ stay unchanged, where $[\mathcal{U}]$ runs over all qubits, which are not included in any of the side conditions. For the initial Hamiltonian we can simply separate the sums over the qubits of different side conditions, because they have no overlap.
For $M$ disjoint side conditions one can use the initial Hamiltonian
\begin{equation}
	H_{\textrm{init}} = \sum_i^{[\mathcal{U}]} \varepsilon_i \tilde{\sigma}^{(i)}_{z} + \sum_i^{[\mathcal{C}_1]} \eta^{(1)}_i \tilde{\sigma}^{(i)}_{z} + ... + \sum_i^{[\mathcal{C}_M]} \eta^{(M)}_i \tilde{\sigma}^{(i)}_{z},
	\label{eq:Hinit_multiSC}
\end{equation}
where the sum $[\mathcal{C}_k]$ runs over all qubits that are included in the $k$-th side condition.
The coefficients $\eta^{(j)}_i$ are chosen such that the side condition $j$ with target  $c_{j}$ is satisfied, i.e., for each of the $M$ side conditions with $\sum_i^{[\mathcal{C}_j]}\tilde{\sigma}^{(i)}_{z} = +c_j$, the local fields are chosen to fulfill $\sum_i^{[\mathcal{C}_j]}\eta^{(j)}_i=-\langle c_j \rangle$. 
Furthermore, the exchange Hamiltonian can be written as
\begin{align}
\label{eq:Hexchange_multipleDisjoint}
	H_{\textrm{exch.}} &= \gamma_e \left(\sum_{\langle i,j \rangle}^{[\mathcal{C}_1]^*} \tilde{\sigma}^{(i)}_+ \tilde{\sigma}^{(j)}_-  +  h.c. \right. \\ \nonumber
	&+ ... \\ \nonumber
	&+ \left. \sum_{\langle i,j \rangle}^{[\mathcal{C}_M]^*} \tilde{\sigma}^{(i)}_+ \tilde{\sigma}^{(j)}_-  +  h.c.
	\right),
\end{align}
where $[\mathcal{C}_k]^*$ denotes the sum over all neighboring pairs of qubits in $[\mathcal{C}_k]$.
Everything else stays unchanged compared to the dynamics for a single side condition.

If one or more side conditions include qubits that cannot be laid out in neighboring positions, we need ancilla qubits and additional parity constraints to realize a parity layout in which all qubits which are included in the same side condition are adjacent. If we consider the example in the main text with an additionally side condition like $c_2=\sigma_{z}^{(3)}\sigma_{z}^{(5)}+\sigma_{z}^{(4)}\sigma_{z}^{(5)}+\sigma_{z}^{(5)}\sigma_{z}^{(6)}$ we do not need any ancilla qubits.
Possible conflicts between side conditions and parity constraints are further discussed in Appendix~\ref{appendix_InterplayParityConstraintsSideConditions}.

\section{Multiple level crossings
\label{appendix_multilevelcrossing}}
\begin{figure}[t]
	\centering
\includegraphics[width=\columnwidth]{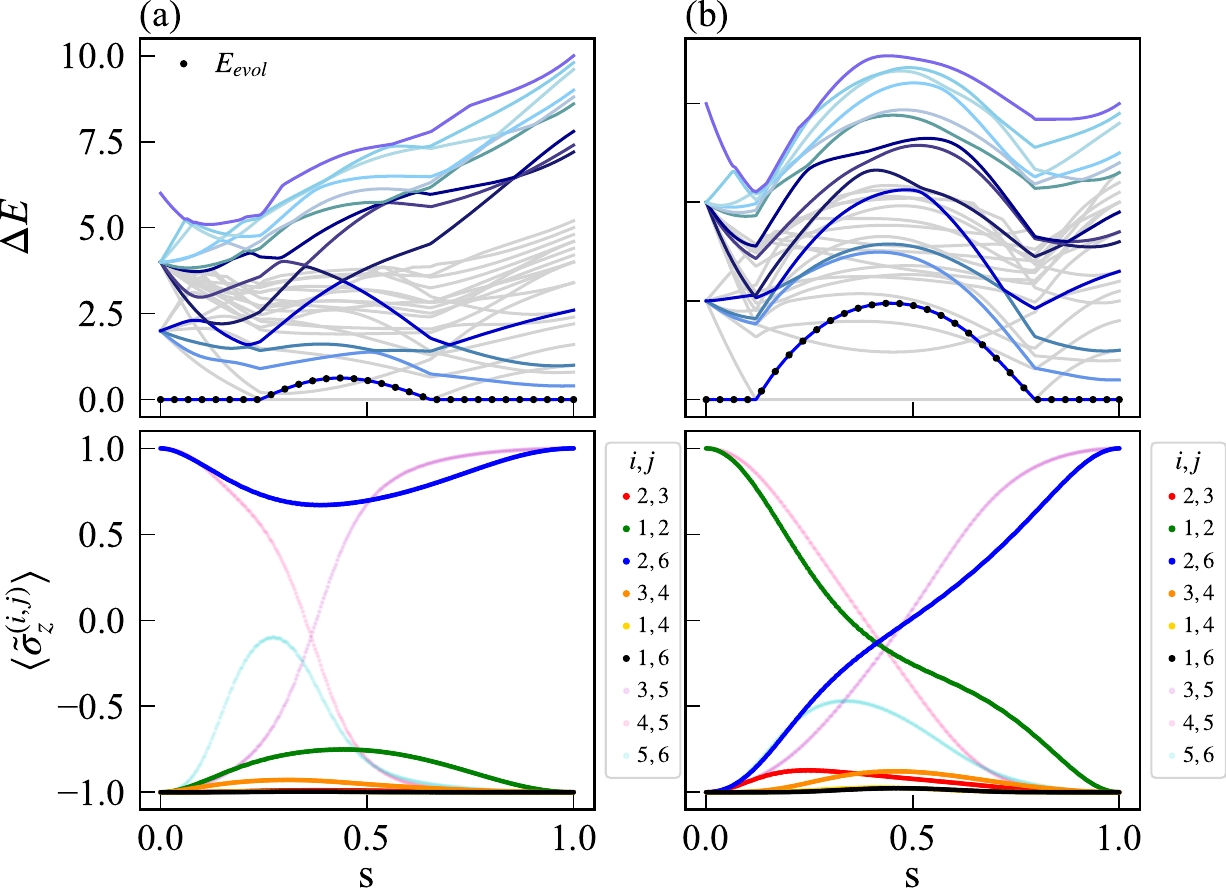}
\caption{%
Adiabatic sweep for a system with $N=9$ qubits. 
The upper panel shows the instantaneous energy spectrum, where all instantaneous eigenenergies $E_i$ of the time-dependent Hamiltonian $H(t)$ [see Eq.~\eqref{eq:Ht}] are given with respect to the ground state energy $E_0$, i.e., ${\Delta E = E_i - E_0}$.
As in Fig.~\ref{fig2}, the colored lines correspond to the energies of the subspace of eigenstates of $H(t)$ satisfying the side conditions. The kinks are level crossings between the energies $E_0$ and $E_1$ in the spectrum. At the level crossing two subspaces meet, that of the conditioned Hamiltonian and that of the full Hamiltonian.
The Figures in the bottom row show the corresponding time-dependent expectation values of all expectation values $\langle\tilde{\sigma}_{z}^{(i,j)} (t)\rangle$ for each site $(i, j)$ of the system.
The side condition is given by
${
-4 = \langle\tilde{\sigma}_{z}^{(2,3)} + \tilde{\sigma}_{z}^{(1,2)} + \tilde{\sigma}_{z}^{(2,6)} + \tilde{\sigma}_{z}^{(3,4)} + \tilde{\sigma}_{z}^{(1,4)} + \tilde{\sigma}_{z}^{(1,6)}\rangle
}$
\textbf{(a)} The initial state is given by
${\Psi_{(a)} =
\ket{-1~ -1~ ~1~ -1~ -1~ -1~ -1~ ~1~ -1}}$.
The time evolution of the system state energy $E_{\rm{evol}}$ crosses the energy levels of the first and second excited states. 
\textbf{(b)} The initial state is given by
${\Psi_{(b)} = \ket{-1~ 1~ -1~ -1~ -1~ -1~ -1~ 1~ -1}}$.
The time evolution of the system state energy $E_{\rm{evol}}$
crosses the energy levels of the first, second and third excited states and ends up in the ground state level.
For both cases, (a) and (b), the initial state and the unconditioned ground state satisfy the side condition.
}
\label{fig10}
\end{figure}

The combination of exchange terms and spin-flip terms may lead to \textit{multiple level crossings} in the instantaneous energy spectrum even if the conditioned ground state is already the ground state of the optimization problem without side conditions.
We call it \textit{multiple level crossings}, when the instantaneous conditioned ground state energy (lowest blue line in Fig.~\ref{fig10}) crosses multiple instantaneous energy levels, which do not lie in the subspace of side-condition-satisfying states (grey lines in Fig.~\ref{fig10}).
Fig.~\ref{fig10}  depicts the same random example as we used in the main text [see Eq.~\eqref{loc_fields_main} and Fig.~\ref{fig1}{\color{quantumviolet}(b)}], i.e., the same local fields and the same parity layout for the optimization problem. Different initial states for different side conditions are used, which all are satisfied by the unconditioned ground state, such that ${\cgs = \ugs}$.
The first 10 eigenstates of the unconditioned problem Hamiltonian represented as spin configuration states ${\ket{\psi_n} = \ket{n_{23}~ n_{12}~ n_{26}~ n_{34}~ n_{14}~ n_{16} ~ n_{35}~ n_{45}~ n_{56}}}$ with ${n_i \in \{-1, 1\}}$ are
\begin{equation}
\label{compiledSCExample}
{\arraycolsep=1.6pt\def\arraystretch{1}
\begin{array}{rrrrrrrrrrrr} 
\ket{\psi_{0}} &= | &-1 &-1 &1 &-1 &-1 &-1 &1 &-1 &-1 &\left. \right>,
\\
\ket{\psi_{1}} &= | &-1 &-1 &-1 &-1 &-1 &1 &-1 &1 &-1 &\rangle  ,\\
\ket{\psi_{2}} &= | &1 &1 &-1 &-1 &-1 &-1 &1 &-1 &-1 &\rangle  ,\\
\ket{\psi_{3}} &= | &-1 &1 &-1 &1 &-1 &-1 &-1 &-1 &-1 &\rangle  ,\\
\ket{\psi_{4}} &= | &-1 &-1 &-1 &-1 &-1 &1 &1 &-1 &1 &\rangle  ,\\
\ket{\psi_{5}} &= | &-1 &1 &-1 &-1 &1 &-1 &-1 &1 &-1 &\rangle  ,\\
\ket{\psi_{6}} &= | &-1 &-1 &-1 &-1 &-1 &-1 &1 &-1 &-1 &\rangle , \\
\ket{\psi_{7}} &= | &-1 &1 &-1 &-1 &1 &-1 &1 &-1 &1 &\rangle  ,\\
\ket{\psi_{8}} &= | &1 &-1 &-1 &-1 &1 &1 &1 &-1 &-1 &\rangle  ,\\
\ket{\psi_{9}} &= | &-1 &-1 &-1 &1 &1 &1 &-1 &-1 &-1 &\rangle ,
\end{array}
}
\end{equation}
where the corresponding eigenenergies are
${\varepsilon_0 = -4.1}$,
${\varepsilon_1 = -3.7}$,
${\varepsilon_2 = -3.3}$,
${\varepsilon_3 = -3.1}$,
${\varepsilon_4 = -3.1}$,
${\varepsilon_5 = -2.5}$,
${\varepsilon_6 = -2.1}$,
${\varepsilon_7 = -1.9}$,
${\varepsilon_8 = -1.7}$, and
${\varepsilon_9 = -1.5}$.

The exchange term enforces conservation of the sum of all spins included in the side condition, making the system stay in the subspace of side-condition-satisfied eigenstates during time-evolution. 
As the instantaneous ground state does not necessarily lie in this subspace, multiple crossings of the instantaneous energy levels can occur.
This means, that if we only consider the instantaneous spectrum of this subspace, we have an energy spectrum without any crossings between the conditioned ground state and the next excited conditioned state in this subspace.

\section{Adiabatic condition \label{appendix_diabatic}}
\begin{figure}[t]
	\centering 
\includegraphics[width=\columnwidth]{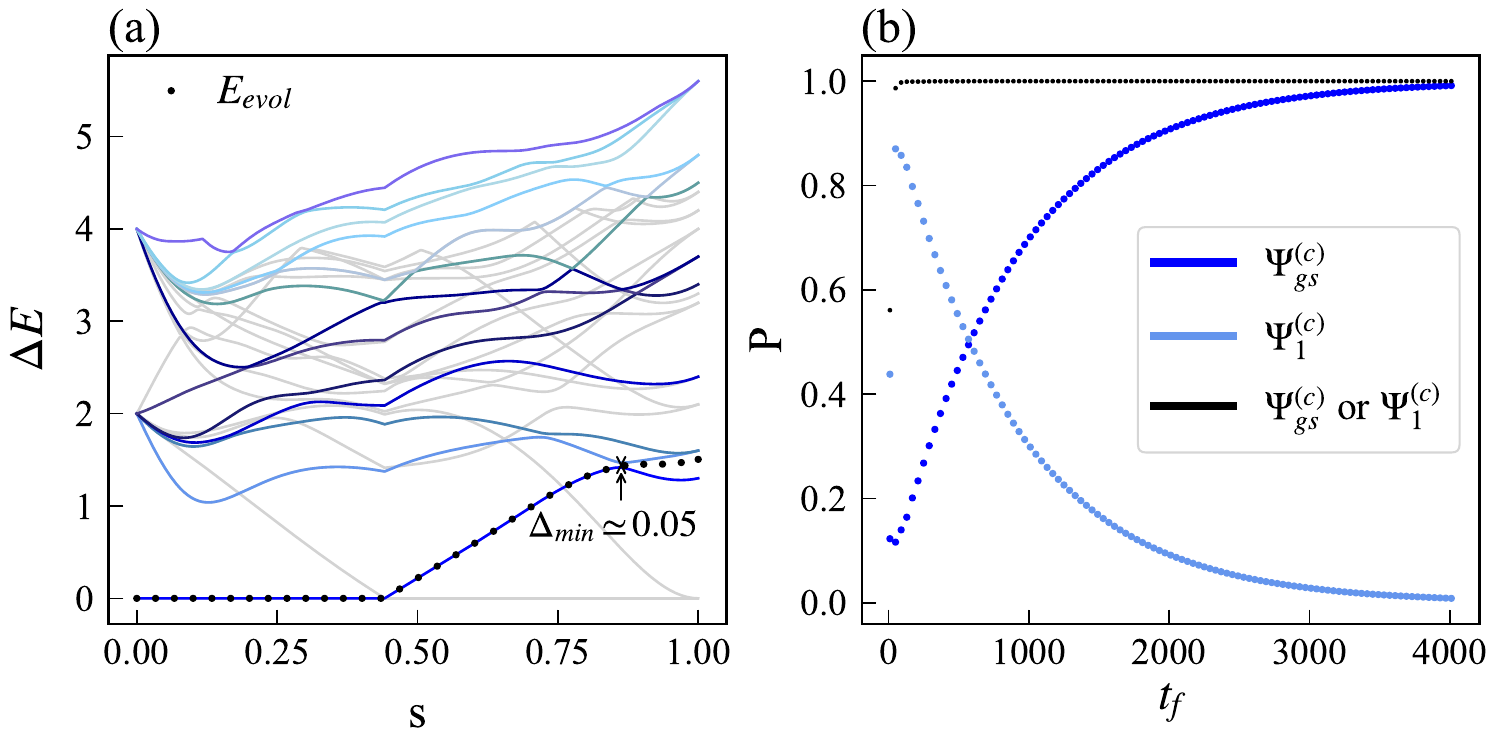}
\caption{%
\textbf{(a)} Non-adiabatic sweep for ${s=t/t_f}$ with ${t_f = 310}$ and ${\Gamma = 4}$. The system leaves the adiabatic path and the process becomes non-adiabatic at the minimum energy gap $\Delta_{\rm{min}}$ in the subspace of conditioned eigenstates (colored lines).
The colored lines are the energies of the instantaneous conditioned eigenstates of $H(s)$ [see Eq.~\eqref{eq:Ht}] which satisfy the side condition, in relation to the instantaneous ground state energy $E_{0}(s)$ of $H(s)$, i.e., ${\Delta E^{c}_{i}(s) = E^{c}_{i}(s) - E_{0}(s)}$.
This figure shows the instantaneous spectrum for a problem instance with a side condition that is satisfied by the second excited state (lowest blue line at $t=t_f$), i.e., the conditioned ground state with ${E^{c}_{0}(s=1) = E_{2} (s=1)}$.
The third excited state of the problem Hamiltonian also satisfies the side condition and is therefore the first excited conditioned state with ${E^{c}_{1} (s=1) = E_{3} (s=1)}$.
Close to $s=0.86$ the annealing process becomes diabatic for ${t_f < 4000}$, because of the small minimum energy gap $\Delta_{\rm{min}}$ between $E^{c}_{0}(s)$ and $E^{c}_{1}(s)$ at ${s\simeq 0.86}$. In this case the system ends up in a superposition of the conditioned ground state (second excited state) and the first excited conditioned state (third excited state).
\textbf{(b)} The probabilities P for the system to be in the conditioned ground state $\cgs$ (${P = |\braket{\Psi(t_f)|\cgs}|^2}$) (dark blue bullets) or in the first excited conditioned state $\Psi^{(c)}_{1}$ (${P = |\braket{\Psi(t_f)|\Psi^{(c)}_{1}}|^2}$) (bright blue bullets) for different final annealing times ${0 < t_f \leq 4010}$. For long enough annealing times (i.e., ${t_f > 4010}$)
the system is in the conditioned ground state at the end of the annealing process. 
}
\label{fig11}
\end{figure}

\begin{figure}[t]
	\centering 
\includegraphics[width=\columnwidth]{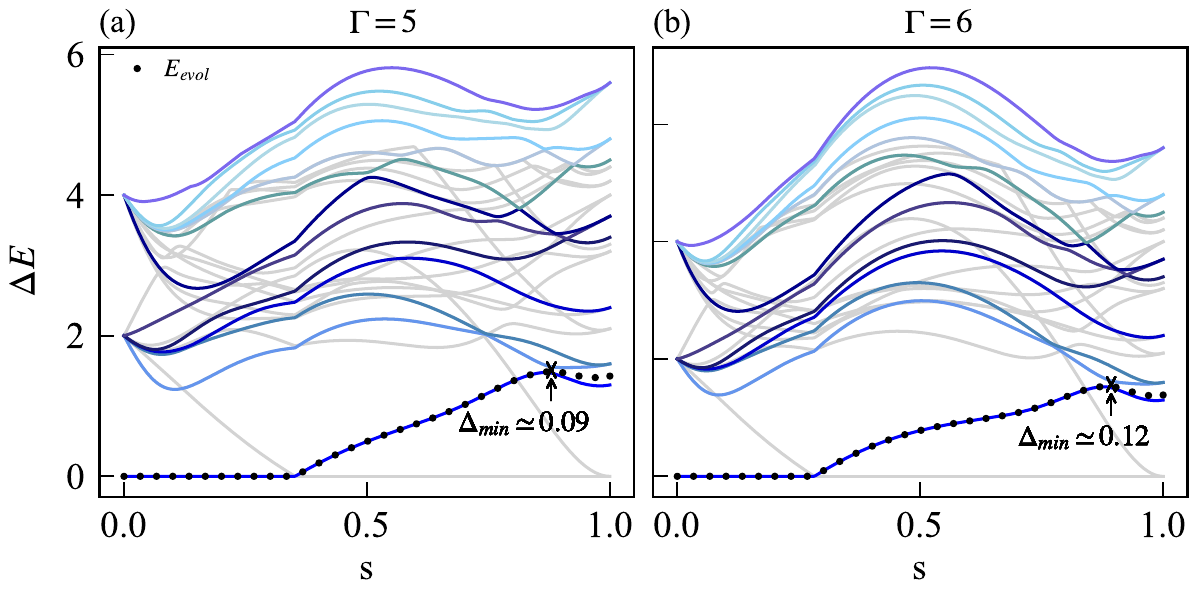}
\caption{Instantaneous spectrum for the same example as given in Fig.~\ref{fig11}{\color{quantumviolet}(a)} but for an annealing protocol with different values for $\Gamma$.
The final annealing time ${t_f=310}$ is chosen here to be the same as in Fig.~\ref{fig11}.
\textbf{(a)} Annealing protocol for ${\Gamma=5}$.
The minimum energy gap is increased to $\Delta_{\rm{min}} \simeq 0.09$ at $s \simeq 0.88$.
\textbf{(b)} Annealing protocol for ${\Gamma=6}$.
The minimum energy gap is increased to $\Delta_{\rm{min}} \simeq 0.12$ at $s \simeq 0.89$.
For higher values of $\Gamma$ the minimum energy gap does not increase much more and moves towards $s=1$.
}
\label{fig12}
\end{figure}

Here we present an example for a non-adiabatic annealing process for the annealing protocol given in Eq.~\eqref{eq:Hs}. Figure \ref{fig11} depicts a problem example with a small minimum energy gap between the conditioned ground state and the next higher conditioned state at $s \simeq 0.86$ in the annealing process.
The given example is implemented by the same parity layout as shown in Fig.~\ref{fig1}, but with the side condition 
\begin{equation*}
    -2 = \langle\tilde{\sigma}_{z}^{(2,3)} + \tilde{\sigma}_{z}^{(1,2)} + \tilde{\sigma}_{z}^{(2,6)} + \tilde{\sigma}_{z}^{(3,4)}\rangle,
\end{equation*}
the local fields ${J_{23}=0.8}$, ${J_{12}=0.6}$, ${J_{26}=1}$, ${J_{34}=0.7}$, ${J_{14}=0.8}$, ${J_{16}=0.6}$, ${J_{35}=0.95}$, ${J_{45}=0.7}$ and ${J_{56}=0.8}$ and the exchange Hamiltonian
\begin{equation*}
    H_{\textrm{exch.}} = \tilde{\sigma}_{+}^{(2,3)} \tilde{\sigma}_{-}^{(1,2)}
+ \tilde{\sigma}_{+}^{(2,3)} \tilde{\sigma}_{-}^{(3,4)}
+ \tilde{\sigma}_{+}^{(1,2)} \tilde{\sigma}_{-}^{(2,6)} + h.c.
\end{equation*}
We consider the minimum energy gap $\Delta_{\rm{min}}$ in the subspace of the conditioned states, i.e., the subspace which is spanned by all eigenstates of $H(s)$ [see Eq.~\eqref{eq:Ht}] satisfying the side condition (colored lines in Fig.~\ref{fig11}).
We denote the energies of the subspace of side-condition-satisfying eigenstates as $E^{c}_{i}(s)$, while the energies of the full instantaneous spectrum are denoted as $E_{i}$.
The conditioned ground state energy and the first excited conditioned state energies are given by $E^{c}_{0}(s)$ and $E^{c}_{1}(s)$, respectively.
To estimate the computation time $t_f$ for this annealing protocol we have to consider the minimum energy gap between $E^{c}_{0}(s)$ and $E^{c}_{1}(s)$, i.e., ${\Delta_{\rm{min}} = \min_{0\leq s \leq 1} E^{c}_{1}(s) - E^{c}_{0}(s)}$.
This differs slightly from the annealing protocols where the minimum energy gap ${g_{\rm{min}} = \min_{0\leq s \leq 1} E_{1}(s) - E_{0}(s)}$ of the full instantaneous energy spectrum limits the process velocity. Here, zero energy gaps in the instantaneous energy spectrum during the annealing process are necessary to allow adiabatic level crossings. This results in an adiabatic evolution inside the subspace of the conditioned states.
By fine-tuning the value of $\Gamma$ one can increase the minimum gap $\Delta_{\rm{min}}$ and therefore reduce the annealing time $t_f$ required for adiabaticity.
In Fig.~\ref{fig12} the same example as in Fig.~\ref{fig11} is given, but for annealing processes with higher values for $\Gamma$. By increasing $\Gamma$ we are able to shift the required final annealing time to lower values.
One can try to minimize computation times, by optimizing the value of $\Gamma$. But this has to be done carefully. If $\Gamma$ becomes much larger (typically quadratically larger or more) compared to the energy scales of the system without $H_{\textrm{exch.}}$, it introduces a new energy scale. In practice, this is critical with respect to the programmable interactions and can worsen error rates.
We remark at this point that the choice of the initial state also affects the size of the minimum gap. 

\section{Degeneracy \label{appendix_degeneracy}}
\begin{figure}[ht]
	\centering 
\includegraphics[width=\columnwidth]{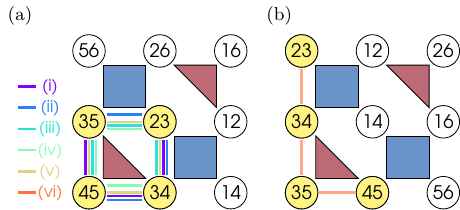}
\caption{Two different parity layouts for the optimization problem given by the Hamiltonian in Eq.~\eqref{eq:logical2bodyH} and the side condition $-2=\langle\tilde{\sigma}_{z}^{(2,3)} + \tilde{\sigma}_{z}^{(3,5)} + \tilde{\sigma}_{z}^{(4,5)} + \tilde{\sigma}_{z}^{(3,4)}\rangle$.
\textbf{(a)} Parity encoding for which the conditioned qubits are layed out  in a loop, such that we are able to find in total five different ways to construct the exchange Hamiltonian. The couplings for the five different possibilities of an exchange Hamiltonian are shown as colored lines between conditioned qubits.
\textbf{(b)} A second parity layout in which the conditioned qubits are adjacent. In this layout only one valid exchange Hamiltonian exists (salmon colored line).
The colors of the six different coupling lines correspond to the colors used for the numerical results in Fig.~\ref{fig14}.
}
\label{fig13}
\end{figure}

Here we present an example to illustrate how the choice of exchange terms and the parity layout influences the probability distribution in a degenerate ground state manifold for the conditioned problem.
We use a problem example given by the logical graph in Fig.~\ref{fig1} and encoded by the local field vector
${
J = (1.0, -0.5, 1.0, -1.0, 0.5, -0.5, -1.0, -1.0, 1.0)^T}
= {(J_{23}, J_{12}, J_{26}, J_{34}, J_{14}, J_{16}, J_{35}, J_{45}, J_{56})^T}
$.
The first nine eigenstates of the unconditioned problem Hamiltonian represented as spin configuration sates ${\ket{\psi_n} = \ket{n_{23}~ n_{12}~ n_{26}~ n_{34}~ n_{14}~ n_{16} ~ n_{35}~ n_{45}~ n_{56}}}$ with $n_i \in \{-1, 1\}$ are
\begin{equation}
\label{DegeneracyExample2}
{\arraycolsep=1.6pt\def\arraystretch{1}
\begin{array}{rrrrrrrrrrrrr} 
\ket{\psi_{0}} &= | &-1 &1 &1 &1 &-1 &1 &1 &1 &-1 &\rangle , \\
\ket{\psi_{1}} &= | &1 &-1 &-1 &1 &-1 &1 &1 &1 &-1 &\rangle , \\
\ket{\psi_{2}} &= | &-1 &1 &-1 &1 &-1 &-1 &1 &1 &1 &\rangle , \\
\ket{\psi_{3}} &= | &1 &1 &-1 &1 &1 &-1 &1 &1 &-1
&\rangle , \\
\ket{\psi_{4}} &= | &-1 &1 &-1 &1 &-1 &1 &1 &1 &-1 &\rangle , \\
\ket{\psi_{5}} &= | &-1 &-1 &-1 &1 &1 &1 &1 &1 &1
&\rangle , \\
\ket{\psi_{6}} &= | &-1 &1 &-1 &1 &-1 &-1 &-1 &-1 &-1 &\rangle , \\
\ket{\psi_{7}} &= | &-1 &-1 &1 &1 &1 &-1 &1 &1 &-1 &\rangle , \\
\ket{\psi_{8}} &= | &-1 &-1 &-1 &-1 &-1 &1 &-1 &1 &-1 &\rangle , \\
\end{array}
}
\end{equation}
where the corresponding eigenenergies are
$\varepsilon_0 = -5.5$,
$\varepsilon_1 = -4.5$,
$\varepsilon_2 = -4.5$,
$\varepsilon_3 = -3.5$,
$\varepsilon_4 = -3.5$,
$\varepsilon_5 = -3.5$,
$\varepsilon_6 = -2.5$,
$\varepsilon_7 = -2.5$, and
$\varepsilon_8 = -2.5$.
With the side condition 
\begin{equation*}
-2=\langle\tilde{\sigma}_{z}^{(2,3)} + \tilde{\sigma}_{z}^{(3,5)} + \tilde{\sigma}_{z}^{(4,5)} + \tilde{\sigma}_{z}^{(3,4)}\rangle,
\end{equation*}
the conditioned ground state is two-fold degenerated, because state $\ket{\psi_6}$ and $\ket{\psi_8}$ have the same energy and are the lowest eigenstates satisfying the side condition.
If we use the parity layout given in Fig.~{\ref{fig13}\color{quantumviolet}(a)}, we have five ways of constructing a valid exchange term $H_{\textrm{exch.}}$~\eqref{eq:Hexchange}.
\begin{enumerate}
\item[(i)]
\small${
\tilde{\sigma}_{+}^{(2,3)} \tilde{\sigma}_{-}^{(3,4)}
+ \tilde{\sigma}_{+}^{(3,4)} \tilde{\sigma}_{-}^{(4,5)}
+ \tilde{\sigma}_{+}^{(3,5)} \tilde{\sigma}_{-}^{(4,5)}
+ h.c.
}$
\normalsize
\item[(ii)]
\small
${
\tilde{\sigma}_{+}^{(2,3)} \tilde{\sigma}_{-}^{(3,4)}
+ \tilde{\sigma}_{+}^{(3,4)} \tilde{\sigma}_{-}^{(4,5)}
+ \tilde{\sigma}_{+}^{(2,3)} \tilde{\sigma}_{-}^{(3,5)}
+ h.c.
}$
\normalsize
\item[(iii)]
\small
${
\tilde{\sigma}_{+}^{(2,3)} \tilde{\sigma}_{-}^{(3,4)}
+ \tilde{\sigma}_{+}^{(2,3)} \tilde{\sigma}_{-}^{(3,5)}
+ \tilde{\sigma}_{+}^{(3,5)} \tilde{\sigma}_{-}^{(4,5)}
+ h.c.
}$
\normalsize
\item[(iv)] \small
${
\tilde{\sigma}_{+}^{(2,3)} \tilde{\sigma}_{-}^{(3,5)}
+ \tilde{\sigma}_{+}^{(3,4)} \tilde{\sigma}_{-}^{(4,5)}
+ \tilde{\sigma}_{+}^{(3,5)} \tilde{\sigma}_{-}^{(4,5)}
+ h.c.
}$
\item[(v)] \small
${
\tilde{\sigma}_{+}^{(2,3)} \tilde{\sigma}_{-}^{(3,4)}
+ \tilde{\sigma}_{+}^{(3,4)} \tilde{\sigma}_{-}^{(4,5)}
+ \tilde{\sigma}_{+}^{(3,5)} \tilde{\sigma}_{-}^{(4,5)}
+ \tilde{\sigma}_{+}^{(2,3)} \tilde{\sigma}_{-}^{(3,5)}
}$\\
${
+ h.c.
}$
\normalsize
\end{enumerate}
and if we use the parity layout given in Fig.~{\ref{fig13}\color{quantumviolet}(b)} we only have one possibility to construct the exchange Hamiltonian, namely
\begin{equation*}\small
    \label{eq:2ndLayout}
  \text{(vi)}~  H_{\textrm{exch.}} = \tilde{\sigma}_{+}^{(2,3)} \tilde{\sigma}_{-}^{(3,4)}
+ \tilde{\sigma}_{+}^{(3,4)} \tilde{\sigma}_{-}^{(3,5)}
+ \tilde{\sigma}_{+}^{(3,5)} \tilde{\sigma}_{-}^{(4,5)}
+ h.c.
\end{equation*}
For each of the six cases we let the system start in the same initial state ${\ket{\psi_{\rm{init}} (0)} = \ket{-1, 1, -1, -1, -1, -1, 1, -1, -1}}$.
\begin{figure}[t]
	\centering 
\includegraphics[width=0.85\columnwidth]{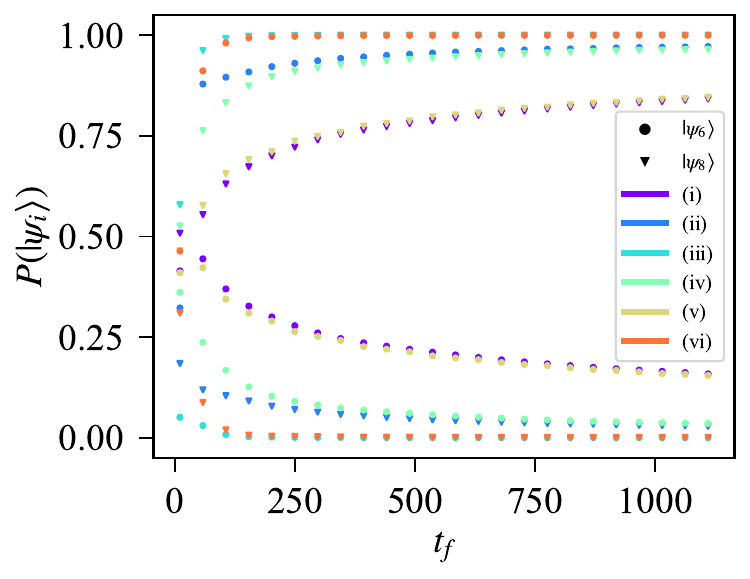}
\caption{
The probability ${P\left( \ket{\psi_i} \right) = |\langle\Psi(t_f)|\psi_i\rangle|^2}$ to find the final system state in the state $\ket{\psi_i}$ is given for the two degenerate ground states  $\ket{\psi_6}$ and $\ket{\psi_8}$ (bullet and triangle). It is calculated for final annealing times $t_f$ between 0 and 1110 for each of the six valid exchange terms (six different colors).
It is clearly visible that the choice of the exchange term (and parity layout) changes the probabilities of the two degenerated ground state.
For the given example we chose all parity constraint strengths to be equal ${C_{ijk(l)}=C=-4}$.
Case (i) and (v) lead to nearly the same result where state $\ket{\psi_8}$ is preferred.
For Case (iii) and (iv) (state $\ket{\psi_8}$ is preferred) the final probabilities are opposite to the case (vi) and (ii) (state $\ket{\psi_6}$ is preferred).
For the cases (iii) and (vi) the system completely ends up in the states $\ket{\psi_8}$ and $\ket{\psi_6}$, respectively.
While case (ii) and (vi) prefer state $\ket{\psi_6}$, all the other cases prefer state $\ket{\psi_8}$ as solution.
It is remarkable that in case (vi) and (iii) a single state is given at the end of the annealing process, instead of a superposition of both degenerated ground states.
}
\label{fig14}
\end{figure}
The numerical results are presented in Fig.~\ref{fig14}.
It demonstrates how the choice of the exchange term influences  the distribution of the probabilities of the degenerate conditioned ground states.
The origin in this dependency on the chosen exchange terms and parity layouts can be explained by the usage of the Schrieffer-Wolff transformation as shown in Ref.~\cite{siebererlechner2018}. The Schrieffer-Wolff transformation is not restricted on the lowest energy sector and works for each set of eigenstates, which have eigenenergies well separated from the rest of the spectrum~\cite{BRAVYI20112793}.
Here we consider a subspace containing the degenerated conditioned ground states, 
which are well separated from the other eigenstates of the spectrum.
We can write the Hamiltonian in the following form
\begin{equation}
\label{eq:HtSW}
H(t) = \alpha_0(t) H_0 + \beta_0(t)V_0 + \beta_1(t)V_1 = H_0(t) + V(t),
\end{equation}
where $H_0$ is the unperturbed Hamiltonian and is given by
\begin{equation}
    \label{eq:H0SW}
    H_0(t) = \alpha_0(t) H_0 = \frac{t}{t_f}\left(H_{\textrm{final}} + H_{\textrm{C}}\right)
\end{equation}
and $V$ is the perturbative term (for $s\simeq 1$)
\begin{align}
    \label{eq:VSW}
    V(t) &= \beta_0(t)V_0 + \beta_1(t)V_1\\ \nonumber
    &= \left(1-\frac{t}{t_f}\right)^2 ~H_{\textrm{init}}\\ \nonumber
    &+ \Gamma \frac{t}{t_f}\left(1-\frac{t}{t_f}\right)~ \left(H_{\textrm{x}} + H_{\textrm{exch.}} \right).
\end{align}
Now we want to determine the effective Hamiltonian for each of the six cases listed above. Depending on the choice of the exchange Hamiltonian we need third, fourth or fifth order perturbation theory.
For the exchange Hamiltonian of case (i), (ii), (iv) and (v) third order is enough. For case (vi) we need fourth order and for case we even need the effective Hamiltonian in fifth order in perturbation theory.
Up to fourth order in perturbation theory, the explicit expression for the effective Hamiltonian is given in Ref.~\cite{BRAVYI20112793} and for higher orders it is a straightforward iterative procedure to get the effective Hamiltonian.
Here we give the explicit expression for the effective Hamiltonian up to third order in perturbation theory, to give the reader an idea how the effective Hamiltonian differs from the one in Ref.~\cite{siebererlechner2018}, where they investigate only single spin flips and no exchange Hamiltonian in the driver.
To name an example we consider the first order term $H_{\textrm{eff}}^{(1)} = PVP$, which is zero in Ref.~\cite{siebererlechner2018} for a pure single spin flip driver. The projector $P$ maps to the subspace of degenerated conditioned ground states.
In our case, $V$ includes the initial Hamiltonian, which maps each of the states onto itself, such that we have $\bra{\psi_i}PVP\ket{\psi_i}=\beta_0(t)v_{0,i}$ where $v_{0,i}$ is the eigenvalue of $V$ acting on the state $P\ket{\psi_i}$.
For a set of degenerated conditioned ground states the effective Hamiltonian is given as 
\begin{equation}
\label{eq:SWeffH}
    H_{\textrm{eff}} = H_{\textrm{eff}}^{(0)} + H_{\textrm{eff}}^{(1)} + H_{\textrm{eff}}^{(2)} + H_{\textrm{eff}}^{(3)} + ...,
\end{equation}
where the different order terms are given by
\begin{align}
    H_{\textrm{eff}}^{(0)} &= H_0\,P,\\ \nonumber
    H_{\textrm{eff}}^{(1)} &= PVP,\\ \nonumber
    H_{\textrm{eff}}^{(2)} &= PV\frac{Q}{E_0-H_0}VP,\\ \nonumber
    H_{\textrm{eff}}^{(3)} &= P\left(V\frac{Q}{E_0-H_0}\right)^2VP ,\\ \nonumber 
    &- \frac{1}{2} \Big[ PV \left(\frac{Q}{E_0-H_0}\right)^2VPVP ,\\ \nonumber 
    &+ PVPV \left(\frac{Q}{E_0-H_0}\right)^2VP \Big].
\end{align}
Let us consider the given example, with the two degenerated conditioned ground states
\begin{align}
    \label{eq:DEGS}
    \ket{\phi_{1}} &=
    \ket{{\color{blue}-1} ~~~1 ~-1 ~{\color{blue}~~~1} ~-1 ~-1 
    {\color{blue}~-1 ~-1} -1},\\
    \ket{\phi_{2}} &=
    \ket{{\color{blue}-1} -1 ~-1 {\color{blue}~-1} ~-1 ~~~~1 
    {\color{blue}~~-1 ~~~~1} -1},
\end{align}
where the blue colored numbers correspond to the conditioned qubits.
In the instantaneous energy spectrum~\eqref{DegeneracyExample2} it is ${\ket{\phi_{1}}\equiv \ket{\psi_6}}$ and ${\ket{\phi_{2}} \equiv \ket{\psi_8}}$.
The operator $P$ is then given by
\begin{equation}
    \label{operatorP}
    P = \sum_{n=1}^{2} \ket{\phi_{n}}\bra{\phi_{n}}
\end{equation}
and the projector $Q$ is defined as $Q = \mathbb{1} - P$.
The operators $H_0$ and $V$ are given by
\begin{equation}
    \label{eq:H0SWoperator}
    H_0 = H_{\textrm{final}} + H_{\textrm{C}}
\end{equation}
and
\begin{equation}
    \label{eq:VSWoperator}
    V = H_{\textrm{init}} + H_{\textrm{x}} + H_{\textrm{exch.}}.
\end{equation}
The operators $H_{\textrm{final}}$, $H_{\textrm{init}}$ and $H_{\textrm{x}}$ are the same for all six cases and given as
\begin{align}
\label{eq:operatorHfinal}
H_{\textrm{final}}  
    &= J_{12} \tilde{\sigma}_{z}^{(1,2)} + J_{23} \tilde{\sigma}_{z}^{(2,3)} + 
    J_{26} \tilde{\sigma}_{z}^{(2,6)}\\ \nonumber
    &+ J_{14} \tilde{\sigma}_{z}^{(1,4)} + J_{34} \tilde{\sigma}_{z}^{(3,4)} + J_{35} \tilde{\sigma}_{z}^{(3,5)} \\  \nonumber
    &+ J_{16} \tilde{\sigma}_{z}^{(1,6)} + J_{45} \tilde{\sigma}_{z}^{(4,5)} + J_{56} \tilde{\sigma}_{z}^{(5,6)},
\end{align}
\begin{align}
    H_{\textrm{init}} &= \tilde{\sigma}_{z}^{(2,3)}  -\tilde{\sigma}_{z}^{(1,2)}  + 
     \tilde{\sigma}_{z}^{(2,6)}\\ \nonumber
    &+ \tilde{\sigma}_{z}^{(3,4)} + \tilde{\sigma}_{z}^{(1,4)} + \tilde{\sigma}_{z}^{(1,6)} \\ \nonumber
    &- \tilde{\sigma}_{z}^{(3,5)} + \tilde{\sigma}_{z}^{(4,5)} + \tilde{\sigma}_{z}^{(5,6)}
\end{align}
and
\begin{align}
    H_{\textrm{x}} &=  \tilde{\sigma}^{(2,6)}_{x} + \tilde{\sigma}^{(5,6)}_{x}  + \tilde{\sigma}^{(1,4)}_{x}\\ \nonumber
    &+ \tilde{\sigma}^{(1,6)}_{x} + \tilde{\sigma}^{(1,2)}_{x}.
\end{align}
On the other hand, the operators $H_{\textrm{c}}$ and $H_{\textrm{exch.}}$ are case dependent.
For the first parity layout in Fig.~\ref{fig13}{\color{quantumviolet}(a)}
there are the four parity constraints
\begin{align} 
\label{eq:operatorHclayouta}
H^{(a)}_{\textrm{C}} &= 
     C_{1234} \tilde{\sigma}_{z}^{(1,2)}\tilde{\sigma}_{z}^{(2,3)}\tilde{\sigma}_{z}^{(3,4)}\tilde{\sigma}_{z}^{(1,4)}\\  \nonumber
    &+ C_{126} \tilde{\sigma}_{z}^{(1,2)}\tilde{\sigma}_{z}^{(2,6)}\tilde{\sigma}_{z}^{(1,6)}\\  \nonumber
     &+ C_{345} \tilde{\sigma}_{z}^{(3,4)}\tilde{\sigma}_{z}^{(3,5)}\tilde{\sigma}_{z}^{(4,5)}\\  \nonumber
     &+ C_{1456} \tilde{\sigma}_{z}^{(1,4)}\tilde{\sigma}_{z}^{(1,6)}\tilde{\sigma}_{z}^{(5,6)}\tilde{\sigma}_{z}^{(4,5)}
\end{align}
and for the second parity layout in Fig.~\ref{fig13}{\color{quantumviolet}(b)}
there are the four parity constraints
\begin{align}
\label{eq:operatorHclayoutb}
H^{(b)}_{\textrm{C}} &= 
     C_{1234} \tilde{\sigma}_{z}^{(1,2)}\tilde{\sigma}_{z}^{(2,3)}\tilde{\sigma}_{z}^{(3,4)}\tilde{\sigma}_{z}^{(1,4)}\\  \nonumber
    &+ C_{126} \tilde{\sigma}_{z}^{(1,2)}\tilde{\sigma}_{z}^{(2,6)}\tilde{\sigma}_{z}^{(1,6)}\\  \nonumber
     &+ C_{345} \tilde{\sigma}_{z}^{(3,4)}\tilde{\sigma}_{z}^{(3,5)}\tilde{\sigma}_{z}^{(4,5)}\\  \nonumber
     &+ C_{2356} \tilde{\sigma}_{z}^{(1,4)}\tilde{\sigma}_{z}^{(1,6)}\tilde{\sigma}_{z}^{(5,6)}\tilde{\sigma}_{z}^{(4,5)}~.
\end{align}
The two parity layouts differ in the last parity constraint.
Here for simplicity all parity constraints have the same strength value $C_{ijk(l)}=-2$.
Finally, we get an individual $H_{\textrm{exch.}}$ operator for each of the six cases (i)-(vi), which are given above.
Now, one can calculate the effective Hamiltonian, which for the two degenerated states results in a $2\times 2$ matrix for each of the six cases.
The four components are given by
\begin{equation}
    \label{eq:effectivHamiltonianMatrix}
    \left(H_{\textrm{eff,} (k)}\right)_{nm}=\bra{\phi_n} H_{\textrm{eff,} (k)}\ket{\phi_m},
\end{equation}
where $n,m=1,2$ and $k \in \{i, ii, iii, iv, v, vi\}$.
From this matrix it is visible that the biases of the different solution states depend on the choice of the parity layout, the corresponding exchange Hamiltonian and the chosen parity constraints strengths $C_{ijk(l)}$. 
The analytic expressions for $H_\text{eff}$ can be readily calculated using Eq.~\eqref{eq:SWeffH}, for example with {\small{MATHEMATICA}}, but will be omitted here due to their length.
The choice of the initial state gives an additional weight which has an influence on the probabilities for the degenerated states with respect to the final annealing times $t_f$.

\section{QAOA-simulation details \label{appendix_qaoa}}

\paragraph{List of side conditions --}Table \ref{tab:sucos} lists the investigated side conditions for the optimization problem depicted in Fig.~\ref{fig1}{\color{quantumviolet}(b)} with local fields as in Eq.~\eqref{loc_fields_main}. Values of $\langle c \rangle$ for which the target state is not in the ten lowest energy eigenstates are omitted.

\paragraph{Parameter optimization --} The parameter updates in the classical optimization loop of the QAOA are done as follows: First we choose a random set of parameters and determine the value of the cost function $E$ [see Eq.~\eqref{energy_penalty_qaoa}]. At each iteration of the algorithm a randomly chosen parameter is updated by a random value in the interval $[-0.1, 0.1]$. If the cost function $E$ decreases, the update is accepted, else rejected. 
For the values shown in Fig.~\ref{fig:QAOA}, this procedure is initialized 100 times with random parameters and 600 consecutive parameter updates per initialization are made. After the optimization procedure the highest reached probability to find the target state is kept.
\begin{table*}[hbt]
  \centering
  \begin{tabular}{r|c|c}
    Side condition indices $\mathcal{C}$ & Target $\langle c \rangle$ & Target index 
    \\ \hline
    $(1, 2),\;(2, 3),\;(2, 6)$  & -1/+1 & 0/2\\
    $(1, 4),\;(1, 6),\;(3, 4)$  & -1/+1 & 1/8\\
    $(3, 5),\;(4, 5),\;(5, 6)$  & -1/+1 & 0/3+4\\
    $(2, 3),\;(3, 4),\;(3, 5)$  & -1/+1 & 0/2\\
    $(1, 2),\;(1, 4),\;(4, 5)$  & -1/+1 & 1/7\\
    $(1, 6),\;(2, 6),\;(5, 6)$  & -1/+1 & 0/3+4\\
    $(1, 2),\;(1, 4),\;(2, 3),\;(4, 5)$  & -2/0/+2 & 1/2/5\\
    $(1, 2),\;(1, 6),\;(2, 6),\;(5, 6)$  & -2/0 & 0/4\\
    $(1, 2),\;(1, 6),\;(2, 3),\;(2, 6),\;(5, 6)$  & -3/-1 & 0/2\\
    $(1, 2),\;(1, 4),\;(2, 3),\;(2, 6),\;(4, 5)$  & -3/-1/+1 & 0/2/5\\
    $(1, 2),\;(1, 4),\;(1, 6),\;(3, 4),\;(4, 5)$  & -3/-1/+1 & 2/1/5\\
  \end{tabular}
  \caption{List of the side conditions used in the QAOA simulations. For a given row the side condition is defined as $\sum_{(i,j)\in \mathcal{C}}\tilde{\sigma}_z^{(i,j)} = c$, where the target state (lowest-energy state fulfilling the given side condition) is listed as an index corresponding to the states given in Eq.~\eqref{compiledSCExample}. For the target indices denoted with 3+4 the probability $P$ is defined as the sum of the probabilities to be in either of the two states.
  } 
  \label{tab:sucos}
\end{table*}

\begin{figure*}[tb]
	\centering 
\includegraphics[width=1.7\columnwidth]{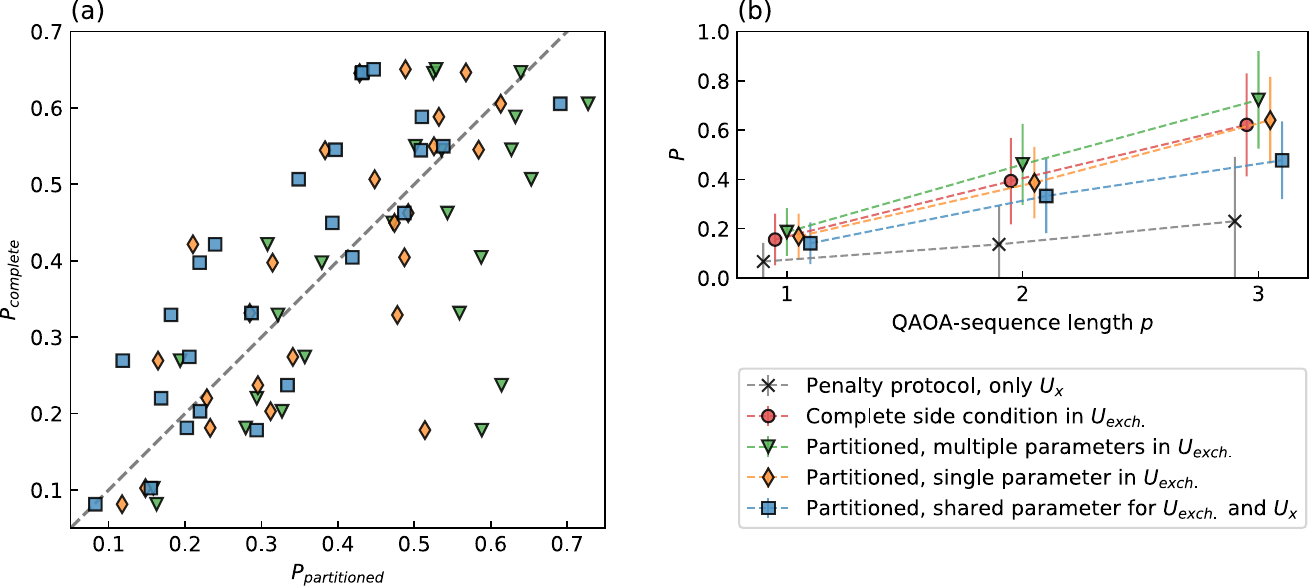}
\caption{\textbf{(a)} Comparison between using the complete side condition vs. different methods of assigning parameters to the partitioned driver Hamiltonian for $p=2$. A point $(P_\textrm{partitioned}, P_\textrm{complete})$ represents the maximal probabilities reached for a specific side condition and value of $c$ for both protocols, respectively.
\textbf{(b)} Mean maximal reached probability $P$ for the QAOA using different mixing unitaries for different sequence-lengths $p$, averaged over all side conditions and values of $\langle c \rangle$. The error bars denote the standard deviation. The points are offset on the x-axis for better visibility.
}
\label{fig15}
\end{figure*}

\paragraph{Partitioning and parametrization of the exchange unitary --\label{appendix_partitioning}}Decomposing the complete unitary $U_\textrm{exch.}(\delta) = \exp(- i \delta H_{\textrm{exch.}})$ into single- and two-qubit gates typically introduces many extra terms, even for exchange terms involving only pairs of qubits. %
%
While such a decomposition is in principle realizable, it is simpler to sequentially apply the different parts of $U_\textrm{exch.}$.
Furthermore, many platforms naturally provide exchange interactions so it can be advantageous to partition the unitaries in a way, such that
parts that share qubits
can be applied to pairs of qubits sequentially.
Note that the `complete' unitary
\begin{equation*}
    U_{\textrm{exch.}}^{\textrm{complete}}(\delta) = e^{- i \delta (\tilde{\sigma}_{+}^{(j)} \tilde{\sigma}_{-}^{(k)}  +  \tilde{\sigma}_{+}^{(k)} \tilde{\sigma}_{-}^{(l)}  + h.c.)}
\end{equation*}
and the `partitioned' unitary
\begin{equation*}
    U_{\textrm{exch.}}^{\textrm{part.}}(\delta) = e^{- i \delta (\tilde{\sigma}_{+}^{(j)} \tilde{\sigma}_{-}^{(k)}  + h.c.)}e^{- i \delta (\tilde{\sigma}_{+}^{(k)} \tilde{\sigma}_{-}^{(l)}  +h.c.)}
\end{equation*}
both commute with the side condition  ${c=\tilde\sigma^{(j)}_{z}+\tilde\sigma^{(k)}_{z}+\tilde\sigma^{(l)}_{z}}$ and are equivalent up to second order in $\delta$. 
After partitioning $U_\textrm{exch.}$, there are multiple ways to parameterize the resulting unitary operators: (i) each of the parts can be assigned an independent parameter, (ii) all parts can have a single parameter or (iii) the parameters for the mixing unitaries on both the conditioned and unconditioned qubits can be kept the same.
The left panel of Fig.~\ref{fig15} shows how the different ways of partitioning perform compared to the complete side condition for all the cases described in Table~\ref{tab:sucos} individually. The right panel of Fig.~\ref{fig15} shows their average performance as a function of the QAOA-sequence length $p$. The probabilities shown are defined as ${P = |\langle \psi|\psi_\textrm{target}\rangle|^2}$, where $\ket{\psi}$ is the variational state prepared by the QAOA and $\ket{\psi_\textrm{target}}$ denotes the lowest-energy state fulfilling the side condition.
For the problem instance investigated here we observe that partitioning with a single parameter performs similar to using the complete side condition, while partitioning with a shared parameter shows a slight decrease in reachable probabilities. Partitioning with multiple parameters shows a slight increase in performance at the cost of additional classical parameters. A more general statement on the performance of the different ways to partition and parameterize the mixing unitary requires further research.
\onecolumngrid
\end{document}